\documentclass[]{emulateapj}
\usepackage{graphicx}

\def\eg{{e.g.}}
\def\ie{{i.e.}}
\long\def\Ignore#1{\relax}
\usepackage{color}
\definecolor{red}{rgb}{0.7,0.1,0.1}

\shorttitle{}
\shortauthors{}

\submitted{Dec 3, 2014}

\begin{document}

\title{Smoothing Rotation Curves and Mass Profiles}

\author{Joel C. Berrier}

\author{J. A. Sellwood}

\affil{Department of Physics and Astronomy, Rutgers the State
  University of New Jersey, 136 Frelinghuysen Road, Piscataway, NJ
  08854, USA}

\begin{abstract}
We show that spiral activity can erase pronounced features in disk
galaxy rotation curves.  We present simulations of growing disks, in
which the added material has a physically motivated distribution, as
well as other examples of physically less realistic accretion.  In all
cases, attempts to create unrealistic rotation curves were
unsuccessful because spiral activity rapidly smoothed away features in
the disk mass profile.  The added material was redistributed radially
by the spiral activity, which was itself provoked by the density
feature.  In the case of a ridge-like feature in the surface density
profile, we show that two unstable spiral modes develop, and the
associated angular momentum changes in horseshoe orbits remove
particles from the ridge and spread them both inwards and outwards.
This process rapidly erases the density feature from the disk.  We
also find that the lack of a feature when transitioning from disk to
halo dominance in the rotation curves of disk galaxies, the so called
``disk-halo conspiracy'', could also be accounted for by this
mechanism.  We do not create perfectly exponential mass profiles in
the disk, but suggest that this mechanism contributes to their
creation.
\end{abstract}



\section{Introduction} \label{sec:intro}

Galaxies at high redshift \citep[\eg][]{Cons11, Newm13} appear far
from smooth, and theory (see \S\ref{sec:galform}) suggests tumultuous
evolution during their early stages of formation.  Yet present-day
galaxy disks have remarkably regular properties and obey quite tight
scaling relations \citep[\eg][]{Tully&Fisher1977}.  In particular,
since they are built from baryonic material that seems most unlikely
to arrive with a smooth distribution of angular momentum, we require
an explanation as to why they have near-exponential light profiles
\citep{Freeman1970} and smooth rotation curves.

Deep photometry \citep[\eg][]{vanderKruit07, EPB08, Martin-Navarro14}
has revealed that minor departures from the exponential light profile
are common \citep[see][for a review]{vdKF11}, but most gradient
changes in the light profile are gentle, and there are no abrupt
discontinuities.

The rotation curves, or radial variation of the circular speed, of
spiral galaxies today also have well-defined regularities
\citep[\eg][for a review]{Sofue&Rubin2001}.  The overall shape of the
rotation curve correlates with the luminosity and circular speed in
the galaxy \citep{Persicetat1996}: slow rises that are almost
continuous to the last measured point characterize low-luminosity
galaxies, a generally flat behavior is found in $\sim L_*$ galaxies,
while a peak at small radii followed by a modest decline before
flattening out is often found in very bright galaxies
\citep[\eg][]{Spekkens06, deBlok08, Salucci2007, Yegorova12}.

However, our present interest is less the overall shape, and more the
general lack of pronounced features in most rotation curves (see also
\S\ref{sec:rctheory}).  \citet{Bahcall&Casertano1985} not only
highlighted this point, and coined the phrase ``disk-halo conspiracy''
to describe the featureless transition from baryon dominated inner
rotation curve (see \S\ref{sec:maxdisk}) to the DM dominated outer
parts, where that can be traced by the 21cm emission line of
\ion{H}{1}.

It should be noted that rotation curves are not completely
featureless.  In particular, \citet{Sancisi2004} stresses that where a
feature is present in the rotation curve it corresponds to a feature
in the observed light profile, or {\it vice-versa}; a phenomenon that
has become known as ``Renzo's rule''.  He discusses six example
galaxies to illustrate the rule.  A further case is the dwarf spiral
galaxy NGC 1560 \citep[\eg][]{Broeils1992, Gentile2010}.  However,
pronounced features such as this are rare, and most cases discussed by
Sancisi required high quality data to be convincing.

In this paper, we suggest a possible explanation for the general
smoothness and lack of pronounced features in the light profiles and
rotation curves of disk galaxies.  We show that any pronounced
features that are introduced into the disk mass profile are quickly
smoothed away by spiral activity, leaving at most mild, and
correlated, variations in both the surface density and circular speed
curve.

\subsection{Galaxy formation} \label{sec:galform}

The disks of galaxies are believed to grow in dark matter (DM) halos
through the settling of gas into a rotationally-supported disk, which
subsequently forms stars on near-circular orbits.  The original
picture, sketched by \citet{WR78}, \citet{FE80}, \citet{Gunn1982} and
others, was that hot gas would cool in the dense halos, but it has
been extended to include cold flows \citep[\eg][]{Keres05, DB06}.  The
current idea is of ``inside out'' disk growth
\citep[\eg][]{Matteucci&Francois1989, Bird2013} since the later
accreted material is expected to have greater angular momentum about
the halo center.

The $\Lambda$CDM paradigm \citep[\eg][]{Springel06}, suggests galaxy
growth is more a story of continuous mass assembly both of the dark
and baryonic matter, punctuated by merger events with a range of
masses, which are believed to have been frequent in the early stages
and to have tailed off towards the present epoch \citep[\eg][]{ZK89,
  Carlberg94, Patton97, Patton02, Conselice03, Lotz06, Lotz08,
  Jogee08, Bertone09, Lotz11}.  The principal plane of the disk is
determined by the net angular momentum vector of the baryons, which
probably slews over time, as does the DM halo shape and mass profile
\citep[\eg]{Katz91, Kazantzidis04, Schneider12, Aumer13}.
Furthermore, current galaxy formation models \citep{Silk03} favor
rapid star-formation for massive galaxies, especially in the early
stages, that preferentially depletes low-angular momentum gas
\citep[\eg][]{Brook11, Kauffmann14, Ubler14}, with wind material being
recycled \citep{OD08, OD10}.  The removal of low-angular momentum
material has the advantage of reducing the expected mass of the inner
disk and bulge, which would otherwise be hard to reconcile with galaxy
mass distributions \citep{vandenBosch01, Sharma05, Dutton2009,
  Sharma2012}.

The steep density profile that the DM halo acquires
\citep[\eg][]{NFW96} as subhalos merge hierarchically is further
compressed by the increasing central attraction from the gas that
cools, settles and forms stars near their centers.  This halo
compression \citep[\eg][]{Blumenthal1986, Gnedin2004,
  Sellwood&McGaugh2005} has been suggested as a means to explain the
disk-halo conspiracy \citep[\eg][]{Burstein&Rubin1985}, although they
suppose that the baryons nowhere dominate the central attraction.

\subsection{Rotation curve from the disk} \label{sec:rctheory}

The connection between the circular speed and the surface density in a
disk is more complicated than for spherical masses \citep{Toomre1963,
  Casertano1983}, because Newton's first theorem \citep[\eg][]{BT08}
does not hold for flattened mass distributions.  In particular, the
central attraction in the plane of a massive, axisymmetric, flat ring
is radially \emph{outwards} over a short radial range just interior to
the ring, and it greatly exceeds the inwards Newtonian attraction of
an equal central mass immediately exterior to the ring.  This behavior
is particularly pronounced in the case of razor-thin rings, and is
only gradually weakened as the ring is thickened vertically.  Thus
ring-like features in the mass distribution of an otherwise smooth,
thin disk can readily give rise to local features in the rotation
curve.  Naturally, the sensitivity of the rotation curve to local
variations in disk surface density becomes less pronounced when the
disk contributes only a fraction of the central attraction, with the
remainder coming from a near-spherical bulge and/or halo.

Features such as ``bumps and wiggles'' have also been noted in some
studies of long-slit rotation curves \citep[\eg][]{Kalnajs1983,
  Kent1986} and of Fabry-Perot velocity maps
\citep[\eg][]{Palunas&Williams2000}. These variations in the circular
velocity profile can often be related to features in the spiral
structure of the disk galaxy being observed.  Streaming motion along
the spiral arms of the galaxy provide a reasonable explanation for the
existence of these features in the rotation curves.

\newpage
\subsection{Disk mass} \label{sec:maxdisk}

A well-observed 1D rotation curve gives an excellent measure of the
total central attraction in the disk mid-plane, but does not contain
enough information to determine the separate contributions of the
baryons and DM \citep[\eg][]{vanAlbada1985}.  The concept of the
maximum disk was introduced by \citet{vanAlbada&Sancisi1986} in order
to place a lower bound on the DM content of galaxy, by giving the
greatest possible mass to the visible matter, limited only by the
requirements that the combined rotation curve nowhere exceeds that
observed and the halo mass distribution should not be hollow.  This
was later refined by \citet{Sackett1997}, who defined a maximum disk
model of a galaxy to be one in which $\ga 85$\% of the circular speed
is due to the stellar disk and bulge, if present, components at
$R=2.2$ disk scale lengths.  This choice of radius is where the
circular speed of an exponential disk reaches its peak value.
Similarly a model has a submaximal disk if the baryon fraction
contributes less at that same radius.

While there is general agreement that the baryons contribute only a
small fraction of the central attraction at most radii in
low-luminosity galaxies \citep[\eg][]{Carignan&Freeman1985} and in
low-surface brightness galaxies \citep[\eg][]{deBlok&McGaugh1997},
there is no consensus as yet on whether large disk galaxies are, or
are not, close to maximal.  For example, \citet{Weiner2001} and
\citet{Bershady2011} reach diametrically opposing conclusions from
different dynamical methods to estimate the stellar mass.  The
above-noted ``Renzo's rule'' is a weak example of a number of indirect
arguments that strongly suggest that disks are more nearly maximal.
Other arguments are as follows:

\citet{Debattista&Sellwood2000} showed that the rotation rates of bars
in submaximal disks are slowed to an extent that is inconsistent with
those observed.  Despite a number of critical papers, no
counter-example has been substantiated to weaken this constraint
\citep[see][for a full discussion]{Sellwood2014, SD14}.

The linear scale for dynamical instabilities in a rotationally
supported disk is \citep{Toom64}
\begin{equation}
\lambda_{\rm crit} = {4\pi^2 G\Sigma \over \kappa^2}.
\end{equation}
For a fixed rotation curve, \ie\ with no change to the radial
variation of the epicycle frequency $\kappa(R)$, this scale varies
with the disk surface density $\Sigma(R)$.  Swing amplification
\citep{GLB65, JT66, Toom81, BT08, Sell13} is strongest where the
wavelength of an $m$-armed disturbance around its corotation circle
\begin{equation}
{2\pi R_{\rm  CR} \over m} \sim 2\lambda_{\rm crit}, \quad\hbox{or}\quad
m \sim {R_{\rm CR} \kappa^2 \over 4\pi G\Sigma}.
\end{equation}
Thus higher disk mass, $\Sigma$, favors lower multiplicity patterns
\citep{Sellwood&Carlberg1984, Athanassoula1987}.  It should be noted
that the patterns in many galaxies are dominated by 2- and 3-fold
symmetry \citep[\eg][]{Davis2014}, indicating heavy disks.

Furthermore, if disk galaxies were indeed submaximal, lower mass to
light ratio, $M/L$, for the stellar component would require an
increased DM fraction within the optical radius, which would increase
the expected scatter in the Tully-Fisher relation between velocity and
luminosity \citep[\eg][]{Zwaan1995, McGaugh2000}.

Thus if galaxy disks are massive, the general smoothness of rotation
curves requires the disk to have a smooth density profile, even though
galaxy formation models (reviewed above) suggest that the infalling
matter is most unlikely to arrive with a smooth distribution of
angular momentum.  It also strengthens the significance of the
``disk-halo conspiracy'', for which no convincing explanation has yet
been proposed.

\subsection{Smoothing mechanism} \label{sec:thmech}

If all material that settles into the disk conserves its detailed
angular momentum, then galaxy formation models would need to be finely
tuned to create the smooth mass profiles that we observe in disks
today.  We noted in \S1.1 that galaxy formation models already require
outflow of preferentially low $L_z$ material.  Material having higher
$L_z$, possibly in the form of cold flows \citep[\eg][]{Keres05, DB06,
  Stewartetal13}, will settle into centrifugal balance at some radius
in the disk determined by its $L_z$.

Past ideas for angular momentum rearrangement to create exponential
profiles include bar formation \citep{Hohl1971, Debattista2006,
  Minchev2012}, a simple analogy with accretion disks
\citep{Lin&Pringle1987, Ferguson&Clarke01}, and scattering off massive
clumps in the disk \citep{Elmegreen2013}.

\citet{Lovelace&Hohlfeld1978} first argued that spiral instabilities
could be responsible for featureless rotation curves.
\citet{Sellwood&Moore1999} noticed, while conducting an unrelated
study, that as a galactic disk grew in their models with a dense
central mass and a rigid halo, the distribution of mass in the disk
was rearranged such that the resulting rotation curve was surprisingly
featureless.  Here, we explore this topic in greater depth using a set
of simulations similar to those used in \citet{Sellwood&Moore1999}
in order to determine the effectiveness of spiral-driven evolution
to smooth out even quite unrealistically non-smooth accreted mass.

\begin{deluxetable*}{cccccccccccccc}
\tablecolumns{13}
\tablecaption{Simulation Information}
\tablehead{
\colhead{Run} & \colhead{Initial }   & \colhead{Final}     & \colhead{Mass} & \colhead{Mean} & \colhead{Annulus} & \colhead{Time} &  \colhead{Step} &
 \colhead{Grid} & \colhead{Size} & \colhead{Disk} & \colhead{Disk}  & \colhead{Bulge} &\colhead{halo}\\
\colhead{ }   & \colhead{particles } & \colhead{particles} & \colhead{added/$\tau_0$ }       & \colhead{Radius} & \colhead{Width}   & \colhead{run}  &  \colhead{size}  
& \colhead{ }   & \colhead{ }    & \colhead{type}  & \colhead{mass} & \colhead{mass} & \colhead{}
}
\startdata\\
 A     & $6$ & $6.7$   & $-3.19$  & $8$  & $4$  & $4000$  & $0.015625$ & $\alpha \times 375$ & $5.0$  & kt    & $1/2$  & $1/2$  & isot(0.5)\\ 
 B     & $6$ & $7.03$  & $-3.37$  & $12$ & $4$  & $16000$ & $0.015625$ & $\alpha \times 225$ & $5.0$  & kt    & $2/3$  & $1/3$  & isot(0.7)\\ 
 C     & $5$ & $7$     & $-3.69$  & $6$  & $2$  & $50000$ & $0.01$     & $\alpha \times 135$ & $5.0$  & kt    & $1/10$ & $9/10$ & isot(0.7)\\ 
 C     &     &         &          & $8$  & $2$  &         &            &                     &        &       & $1/10$ & $9/10$ &          \\ 
 C     &     &         &          & $10$ & $2$  &         &            &                     &        &       & $1/10$ & $9/10$ &          \\ 
 C     &     &         &          & $12$ & $2$  &         &            &                     &        &       & $1/10$ & $9/10$ &          \\ 
 D     & $6$ & $6.7$   & $-3.37$  & $12$ & $2$  & $8000$  & $0.015625$ & $\alpha \times 135$ & $5.0$  & kt    & $2/3$  & $1/3$  & isot(0.7)\\ 
 D     &     &         &          & $10$ & $2$  &         &            &                     &        &       &        &        &          \\ 
 D     &     &         &          & $8$  & $2$  &         &            &                     &        &       &        &        &          \\ 
 D     &     &         &          & $6$  & $2$  &         &            &                     &        &       &        &        &          \\ 
 E     & $6$ & $6.7$   & $-3.37$  & $12$ & $2$  & $8000$  & $0.015625$ & $\alpha \times 135$ & $5.0$  & kt    & $2/3$  & $1/3$  & isot(0.7)\\  
 E     &     &         &          & $10$ & $2$  &         &            &                     &        &       &        &        &          \\ 
 E     &     &         &          & $8$  & $2$  &         &            &                     &        &       &        &        &          \\ 
 E     &     &         &          & $6$  & $2$  &         &            &                     &        &       &        &        &          \\ 
 F     & $6$ & $6.7$   & $-3.18$  & $6$  & $2$  & $5000$  & $0.05$     & $\alpha \times 135$ & $5.0$  & kt    & $2/3$  & $0$    & isot(0.7)\\ 
 G     & $6$ & $6.7$   & $-3.3$   & $4$  & $4$  & $8000$  & $0.04$     & $\alpha \times 135$ & $4.0$  & exp   & $1$    & $0$    & hern(36)) \\
 M     & $5$ & $5.08$  & $-2.88$  & $7$  & $4$  & $250$   & $0.025$    & $\beta  \times 125$ & $15.0$ & mtz   & $1/3$  &        &  $2/3$   \\ 
 N     & $5$ & $5.08$  & $-2.88$  & $7$  & $4$  & $250$   & $0.025$    & $\beta  \times 125$ & $15.0$ & mtz   & $1/3$  &        &  $2/3$   \\
\enddata

\tablecomments{Column 1: Identification for the simulation. Column 2:
  Log of initial number of particles. Column 3: Log of final number of
  particles.  Column 4: Log of fraction of unit mass $M$ added per
  dynamical time $\tau_0$.  Column 5: Location of annulus of added
  particles in system units $a$.  Column 6: Width of the annulus in $a$.  Column 7: Duration of
  the simulation in $\tau_0$.  Column 8: Time step in $\tau_0$.  Column 9:
  Numbers of grid rings, spokes, and planes; $\alpha = 128 \times
  128$, $\beta = 108 \times 128$.  Column 10: Truncation radius of the
  initial disk in $a$. Column 11: Type of initial disk; ``kt'' is the
  Kuzmin-Toomre disk, and ``mtz'' is the Mestel disk.  Column 12:
  Initial disk mass fraction. Column 13: Initial bulge mass fraction.
  A repeated line indicates that the mean radius of annulus was
  changed during the simulation. Column 14: Halo type and parameter;
  ``isot'' is a cored isothermal halo which asymptotes to the value of
  the circular velocity provided in the parentheses, ``hern'' is a
  Hernquist profile with a mass $36$ times that of the disk and a
  scale radius of $10a$, and the $2/3$ halo are rigid potentials of
  the same form as the Mestel disk, with $2/3$ of the total mass.}
\label{table:sims}
\end{deluxetable*}

\section{Methods} \label{sec:methods}

We use $N$-body simulations of a growing disk in order to examine the
effects of spiral activity on the distribution of matter within a disk
and consequently the shape of the rotation curve.  Our models have
both central bulges and extensive halos, and we represent these
spheroidal components as rigid, unresponsive matter that contributes
only to the central attraction experienced by the disk particles.
This approximation has not only reduced the computational effort by a
substantial factor, but it has also enabled us to study the effects of
spiral activity in the disk, which is the major science goal of this
paper, without the complications of changes to the halo density.  We
defer to a subsequent paper a study of the effects of adiabatic
contraction and angular momentum exchange between the disk and live
spheroidal components.

\subsection{Initial disk, bulge and halo}
Most of the models we present start with a Kuzmin-Toomre disk
\citep[][model 1]{Toomre1963} that has the surface density
\begin{equation}
\Sigma(R) = \frac{M q}{2 \pi a^2}\left(1+\frac{R^2}{a^2}\right)^{-3/2},
\end{equation}
where $R$ is the cylindrical radius, $M$ a notional mass, $q$ the disk
mass fraction, and $a$ a length scale.  We truncated the disk at
$R=5a$.  The rigid bulge has the \citet{Hern90} density profile
\begin{equation}
\rho_b(r)=\frac{M_b }{2 \pi b^3}
\left(\frac{r}{b}\right)^{-1}\left(1+\frac{r}{b}\right)^{-3},
\end{equation}
where $r$ is the spherical radius, $M_b$ is the bulge mass, and $b$
is a length scale.  The halo component has a cored isothermal
density profile
\begin{equation}
\rho_h(r) = \frac{V_\infty^2}{4\pi G}\frac{3c^2+r^2}{(c^2+r^2)^2} \ ,
\end{equation}
where $V_\infty$ is the asymptotic circular speed, and $c$ is the core
radius.  For model A (Table~\ref{table:sims}), the disk mass $M_d =
M/2$, the Hernquist bulge has $M_b = M/2$ and $b = a/5$, while the
cored isothermal halo has $V_\infty = 0.5V_0$ (defined in
\S\ref{sec:units}) and $c = 30a$.  The parameters of other models are
given in Table~\ref{table:sims}.

The initial in-plane particle velocities were chosen such that Toomre's
$Q=1.5$, the particles were spread vertically about the mid-plane with
a Gaussian density profile having a standard deviation of $0.1a$, and
they were given vertical velocities derived from the 1D Jeans
equation.

This near-equilibrium initial disk developed a bar in most cases.  As
we here wish to study the effects of spirals only, we allow the bar to
form and settle, and then rearrange the particles by selecting a new
azimuth for each, chosen at random from a uniform distribution, while
preserving all the other phase space coordinates in cylindrical polar
geometry.  This ``shuffling'' procedure, which was also used by
\citet{Hohl1971}, destroys the bar and, because the inner disk is now
dynamically hot, helps to prevent future bar formation in the disk as
the model is evolved.

Our accretion experiments begin with this hot disk, and it is
therefore of no significance to this study what occurred to the disk
prior to the addition of mass.  We simply use this early part of the
evolution to create an initial, axisymmetric, bar-stable disk as a
suitable initial model.

\subsection{Units}
\label{sec:units}
We adopt $M$ as our unit of mass and $a$ as our unit of length.  Thus
our unit of velocity is $V_0 = (GM/a)^{1/2}$ and time unit, or
dynamical time, is $\tau_0 = a / V_0 = (a^3/G M)^{1/2}$.  Henceforth,
we use units such that $G=M=a=1$.  A suitable scaling to physical
units for most of our models is to choose $a = 0.5\;$kpc and $\tau_0 =
1.5\;$Myr, which yields $M \simeq 1.2 \times 10^{10}\;$M$_\odot$ and
$V_0 \simeq 326\;$km/s, but other scalings would be more appropriate
for some of our models.

\subsection{Accretion rules}
\label{sec:rules}
We wish to mimic disk growth in a manner that enables us to control
the distribution of accreted matter.  We therefore add particles to
the disk placing them on circular orbits at a chosen radius and
randomly distributed in azimuth, which crudely mimics the settling of
gas that subsequently forms stars.  This technique captures the single
most important dynamical consequence of gas dissipation, because it
supplies dynamically cool particles that can participate in subsequent
spiral activity, which would otherwise fade as random motion rises
among the older particles.

We employ a wide variety of rules for the addition of fresh material
in order to explore the extent to which spiral activity can rearrange
matter within in the disk.  Therefore, the rules are not always
realistic, and were deliberately chosen to be quite unrealistic in
some cases, in order to determine whether the outcome of disk growth
is, or is not, sensitive to the distribution of angular momentum in
the accreted matter.  In all cases, we have added particles to the
disk in one or more annuli, with various mean radii and widths, and
have experimented with both uniform and Gaussian distributions for the
particles in the annulus.

We have experimented with a variety of accretion rules, by varying the
following parameters:
\begin{itemize}

\item The radial extent of the annulus: it ranged from a delta
  function to one having a width $8a$, with $4a$ being the most often
  used.

\item The mean radius of the annulus, $R_{\rm mean}$ was varied over a
  wide range: $0.5 \leq R_{\rm mean}/a \leq 12$.

\item The rate  of accretion of new particles,  which ranged over $2.0
  \times  10^{-4}$ to  $1.28 \times  10^{-2}$,  in units  of $M$,  the  
  notional mass, per dynamical time. Larger values tended to cause the 
  models to develop bars quickly, or for large numbers of particles to
  leak off the grid, or both.

\item The radial profile of the accretion annulus.  All simulations in
  Table~\ref{table:sims} used radii chosen from a uniform
  distribution.  We also tried experiments with radii selected from a
  Gaussian in $R$, for which the results were quite similar.

\item The mean radius of the annulus was changed over time, in some
  cases.  We experimented both with stepping the mean outwards with
  time, to be consistent with ``inside-out'' growth, and we also tried
  ``outside-in'' growth in a few cases.

\item All the above rules were axisymmetric, but we also tried
  azimuthally localized accretion.  Accretion onto a fixed azimuth as
  the disk rotates underneath, which is more what might be expected,
  leads over time to an almost axisymmetric ring.  However, as an
  extreme example, we rotated the accreting patch at the local
  circular speed to ensure the build-up a strongly non-axisymmetric
  distribution.  (Note that since we suppress the dipole term in the
  field determination, this is effectively adding matter in two
  patches 180$^\circ$ apart.)

\end{itemize}

\begin{figure}
\begin{center}
\includegraphics[width=.9\hsize]{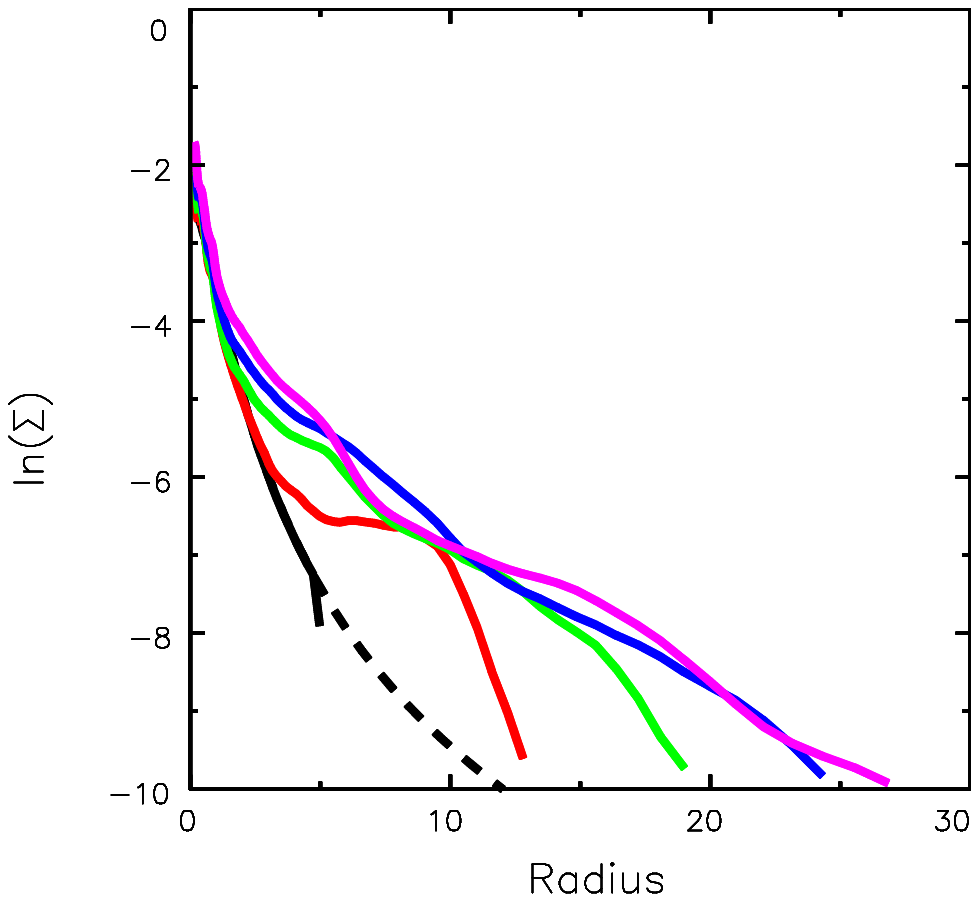}

\medskip
\includegraphics[width=.9\hsize]{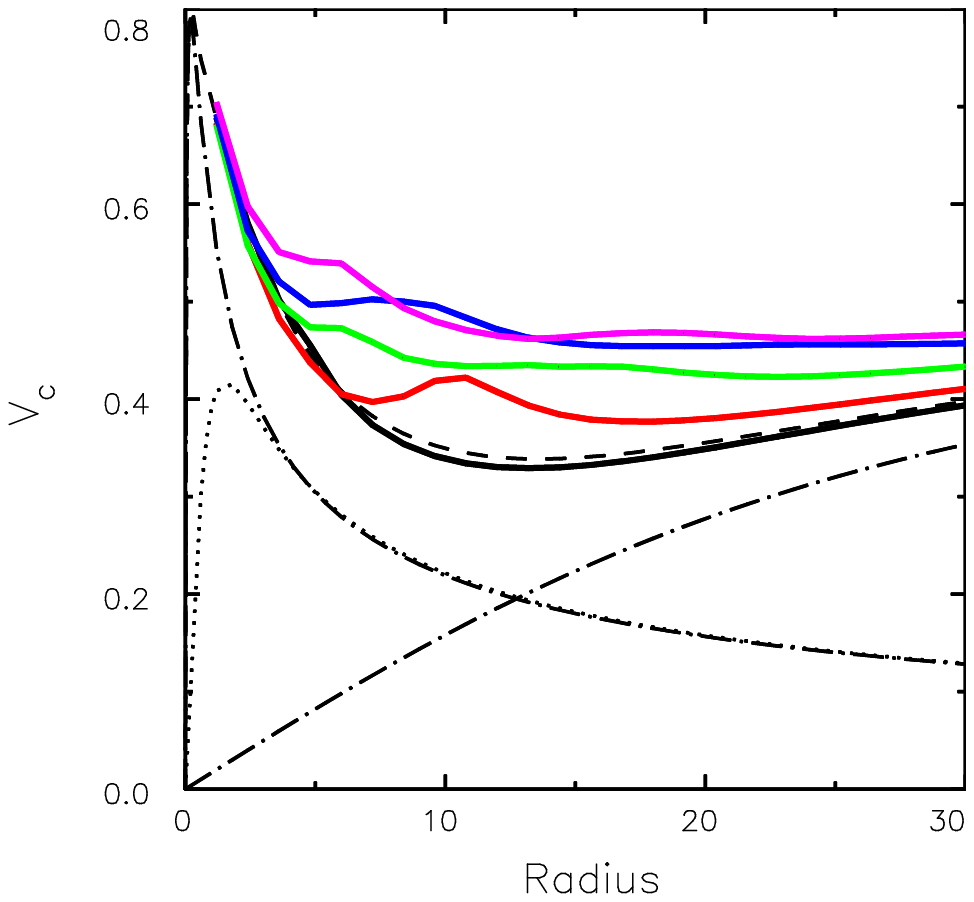}
\end{center}
\caption{The upper panel shows the evolution of the surface density in
  model A (Table~\ref{table:sims}).  The dashed line shows the
  theoretical surface density of the untruncated KT disk, and the
  solid black line the initial density of the particles.  The red,
  green, blue and magenta curves show respectively the surface density
  profiles at intervals of $1000\tau_0$.  The particles in this
  simulation were added in an annulus from $6 \le R/a \le 10$.  The
  circular speed at the same times is shown by the colored lines in
  the lower panel.  The black lines show the initial contributions of
  the disk (dotted), equal mass bulge (decreasing dot-dashed line),
  and halo (rising dot-dashed line), the combined theoretical curve
  (dashed), and the measured circular speed (solid).  }
\label{fig:282}
\end{figure}

\subsection{Numerical details}
We compute the gravitational field of the disk particles using a 3D
polar grid \citep[see][]{Sellwood&Valluri1997}; full details of the
$N$-body code are given in the on-line manual available at
http://www.physics.rutgers.edu/$\sim$sellwood/manual.pdf.  For model
A, the grid had 128 spokes, 128 rings, and 375 planes with vertical
spacing $0.1a$; the in-plane grid was refined to twice these values in
a number of cases.  We determined the gravitational forces using
sectoral harmonics $0 \le m \le 8$ only, with $m=1$ also excluded to
avoid unbalanced forces from the rigid mass components due to possible
asymmetries in the distribution of particles.

We use a cubic-spline softening rule, with a softening length of
$0.1a$, although we doubled this value when we used a grid with
greater vertical spacing and fewer planes.  We employed $10^6$
particles in the initial disk, but the number increased five-fold by
the end.  We adopted a basic time step of $\tau_0/64$, which we
increased by five factors of two in zones of ever larger radii
\citep{Sell85}.  These parameters were for model A; values for
other runs are given in Table~\ref{table:sims}.

\section{Results}   \label{sec:results}

In this section we discuss the results of a few specific simulations,
whose parameters are summarized in Table~\ref{table:sims}.  We
explored numerous other models besides those in
Table~\ref{table:sims}, many of which were minor variants of this
basic set with differing rates of accretion or the width or profile of
the annulus.  The behavior turned out to be insensitive to almost all
accretion parameters, with few noteworthy differences from those
reported here.
\begin{figure}
\begin{center}
\includegraphics[width=.9\hsize]{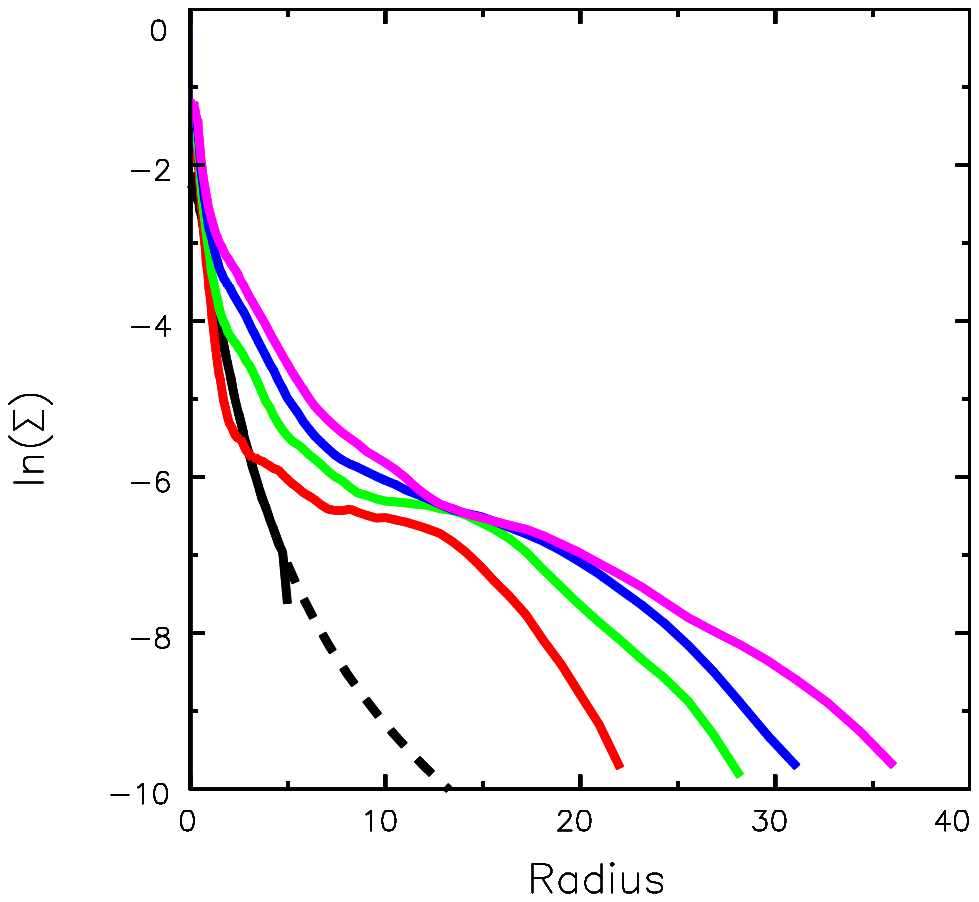}

\medskip
\includegraphics[width=.9\hsize]{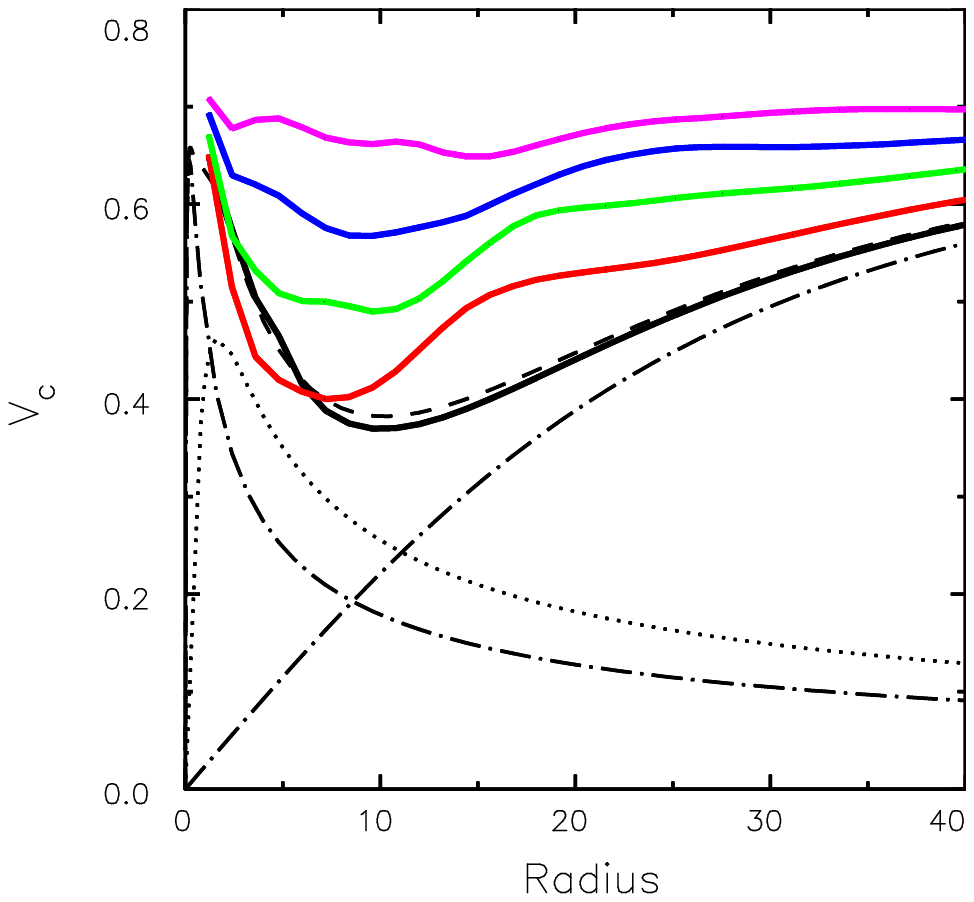}
\end{center}
\caption{Same as for Figure~\ref{fig:282}, but for model B
  (Table~\ref{table:sims}). The time interval between
  curves is $4000\tau_0$.  Note that the final circular velocity,
  magenta, is a reasonably flat rotation curve, with no feature in the
  transition from bulge-disk to halo dominance in the profile.}
\label{fig:307}
\end{figure}

\subsection{Model A}  \label{Disks}

We first describe model A, for which the evolution of the surface
density and rotation curve are illustrated in Figure~\ref{fig:282}.
After the bar had formed and been erased, we began adding particles to
the disk at a rate of $1280$ particles per $\tau_0$, or approximately
$0.128$\% of the initial disk mass, continuously for a period of
$3125\tau_0$.  We placed the particles over the radial range $6 \leq R
\leq 10$, which caused a local shoulder in the surface density at
first, shown by the red line in the upper panel of
Figure~\ref{fig:282}, that subsequently smoothed out as the disk mass
rose, despite the fact that we continued to add mass over the same
radial range.  The disk spread significantly as its mass rose
five-fold, to yield a quite smooth quasi-exponential surface density
profile by the last time shown (magenta line).

The lower panel shows the evolution of the rotation curve, which began
with a pronounced dip between the contributions of the dense bulge and
the initial disk in the inner part and the cored halo farther out.
While the dense bulge continued to maintain high circular speeds in
the center, the broad initial dip is gradually erased.  This is
perhaps not too surprising, since we are increasing the disk surface
density just interior to the middle of the dip, but the final rotation
curve has hardly a feature at all as the substantial added mass has
been spread radially.  However, the bulge still dominated motion in
the center of the model.

\begin{figure}
\begin{center}
\includegraphics[width=.9\hsize]{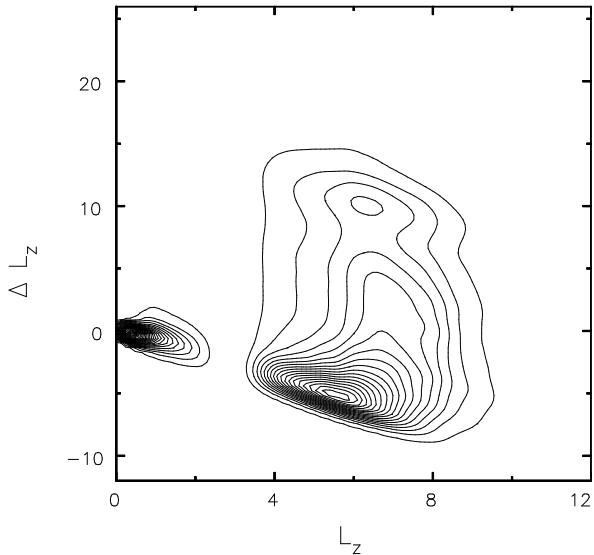}
\end{center}
\caption{The density of particles in model B as a function of their
  initial angular momentum, $L_{z,i}$, and angular momentum change,
  $\Delta L_z$.  The initial particles form the density contours on
  the left, while the nine times more numerous particles that were
  added to the disk form the contours on the right.}
\label{fig:307dLz}
\end{figure}

The evolution of the model was dominated by bi-symmetric spirals, while
a small bar of semi-major axis $\sim 0.5a$ appeared around $t = 700
\tau_0$, although it did not persist.  A larger bar with semi-axis
$\sim 6a$ developed around $t = 3000 \tau_0$, but the effect of the
spirals on the initial disk had caused most of the evolution long
before that bar became dominant.

\begin{figure}
\begin{center}
\includegraphics[width=.9\hsize]{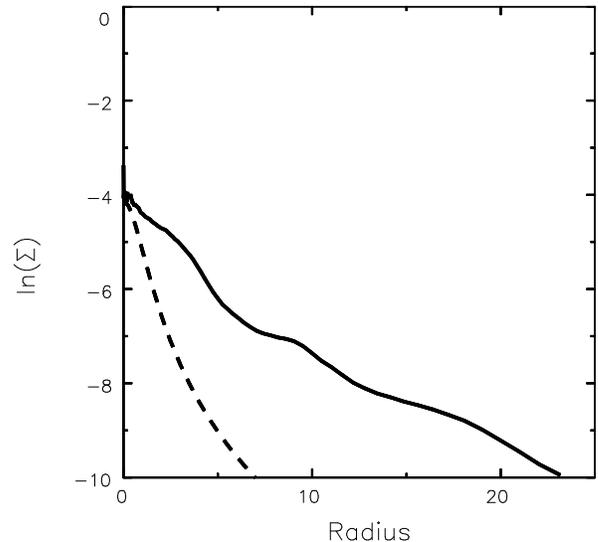}

\medskip
\includegraphics[width=.9\hsize]{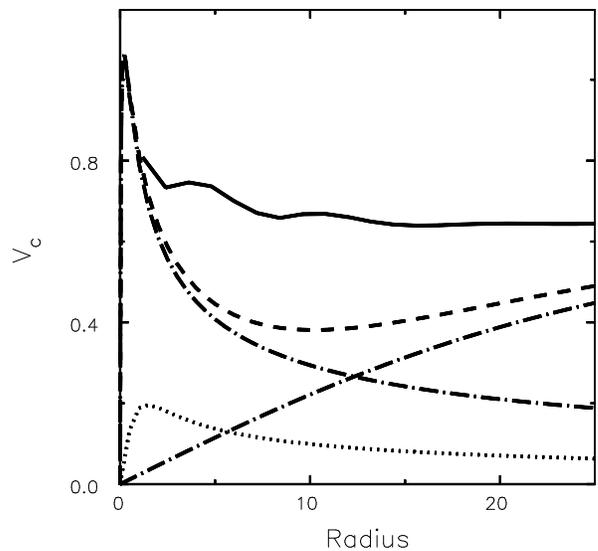}
\end{center}
\caption{Same as for Fig.~\ref{fig:282} but for model C, in which the
  disk grew from 1\% of its final mass.  The solid curves illustrate
  the final disk only.}
\label{fig:305}
\end{figure}

\subsection{Model B}
The evolution of the mass distribution in model B is illustrated in
Figure~\ref{fig:307}.  This model differs from model A in having a
lower mass bulge, the added particles are placed farther out ($10 \leq
R/a \leq 14$) at half the rate, and it was also run for a much
longer time.  The mass of the final disk had increased by more than
ten-fold, and was represented by $>10^7$ particles and the final
rotation curve was smooth and nearly flat.

In this case, the initial evolution was dominated by more multi-arm
spiral patterns, up to $m=6$ at first, but the multiplicity of arms
dwindled to $m=2$ by about half-way through the simulation.  Again, a substantial
bar, with semi-major axis $\sim 15a$, formed by the end of this
simulation, but spiral activity dominated for well over half of the
duration.

Note that the annular range of mass addition was outside the
initial disk, yet the final surface density profile has no gap, which
requires there to have been substantial rearrangement of angular
momentum among and between both the original and added particles.
Figure~\ref{fig:307dLz} contours the density of particles as functions
of $L_{z,i}$ and $\Delta L_z$, where $L_{z,i}$ is the initial angular
momentum at the start, or when the particle was added, and $\Delta
L_z$ is the change in that value by the end of the simulation.  The
original particles are in the blob on the left, while the added
particles have larger $L_{z,i}$, and the densest concentrations lie
close to the line of slope $-1$, suggesting that many particles have
sunk far towards the center, but few have become retrograde.
Obviously, losses of angular momentum must be balanced by gains, as
there is no external torque, and some of the added particles have more
than doubled their initial $L_{z,i}$.  The local maximum in the center
of the panel forms much later in the simulation, around $15000-16000
\tau_0$. This is the same time frame over which the bar seems to grow
significantly.

\begin{figure}
\begin{center}
\includegraphics[width=.9\hsize]{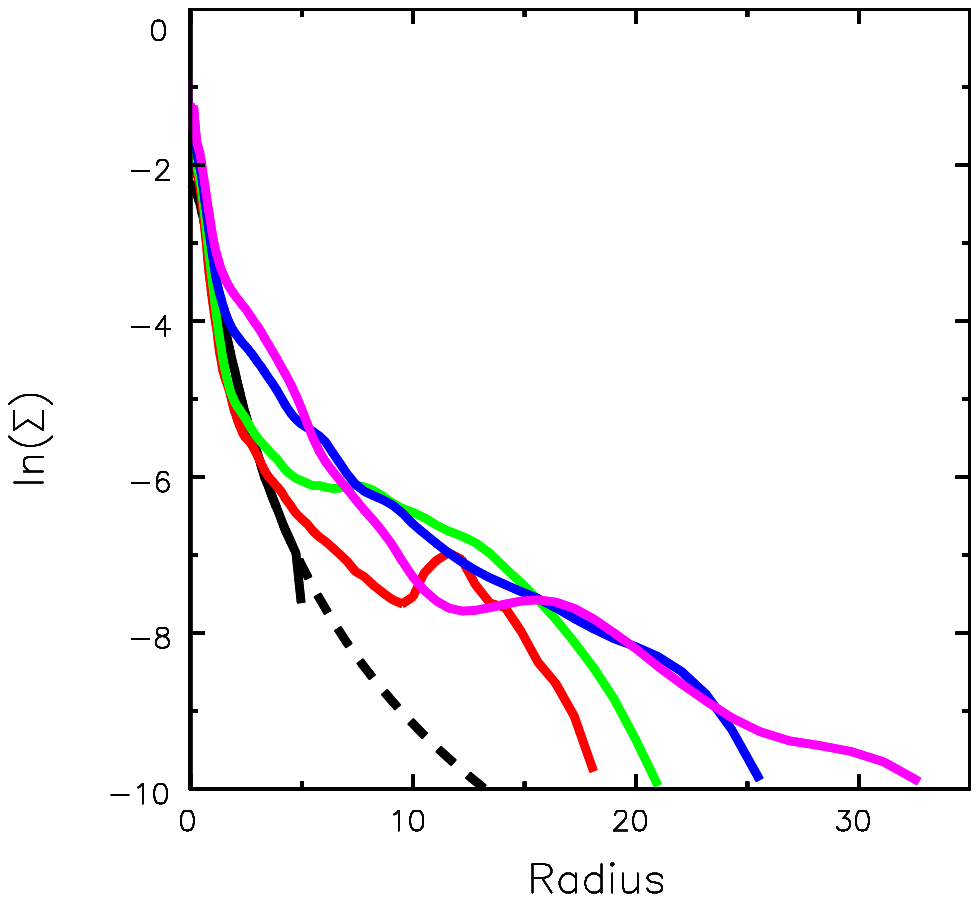}

\medskip
\includegraphics[width=.9\hsize]{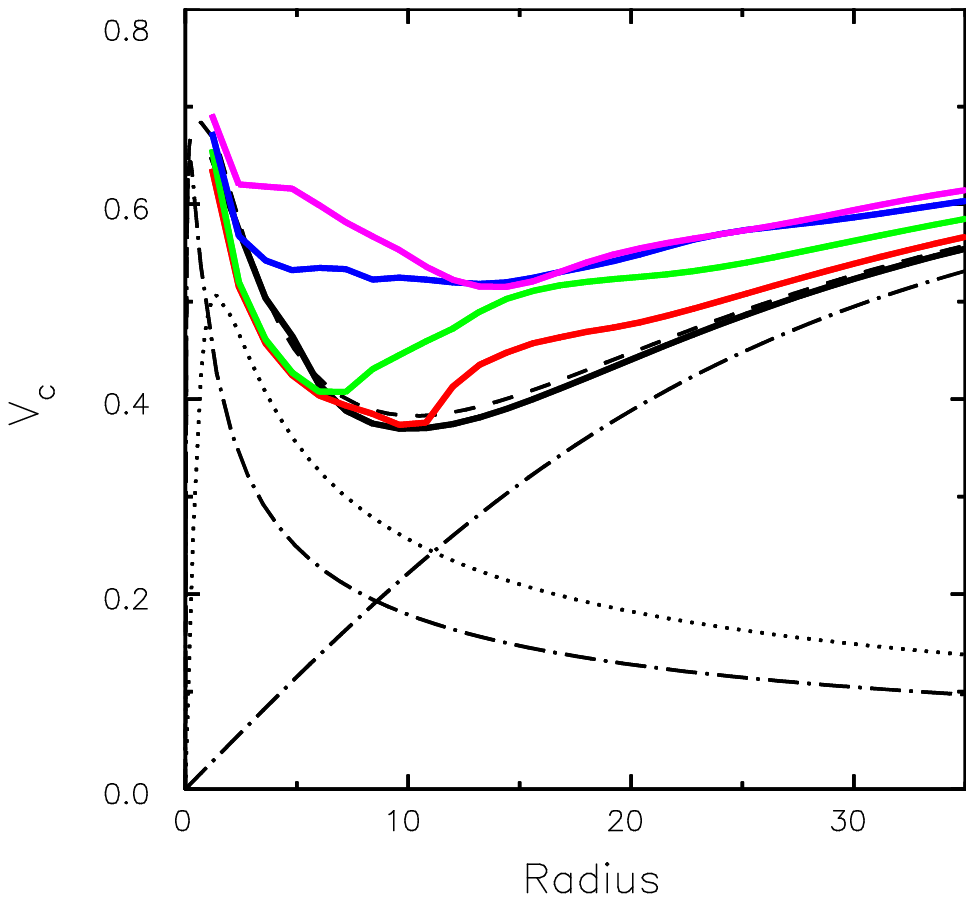}
\end{center}
\caption{ Surface density and rotation curve for sample disk grown
  ``outside-in''. \emph{Upper:} The surface density of the disk at
  intervals of $2000 \tau_0$.  The final surface density (magenta
  line) has a feature near $R=12$, which is the approximate the bar
  semi-major axis late in the simulation.  The black curve is the
  initial disk, while the red, green, blue, and magenta represent the
  later respective times.  \emph{Lower:} The radial variation of the
  circular speed at the same times as the upper panel. }
\label{fig:316}
\end{figure}

\subsection{Model C}
Figure~\ref{fig:305} illustrates the initial and final surface
density, upper panel, and circular velocity, lower panel, for model C.
In this case the disk mass increased 100-fold as the mean radius of
the accreted mass was increased.  The rotation curve is still
dominated by the dense bulge in the inner parts, but becomes
essentially flat outside the bulge region.  The final surface density
shows a decrease around seven scale lengths, the same radius as the
bar that forms.  

\begin{figure}
\begin{center}
\includegraphics[width=.9\hsize]{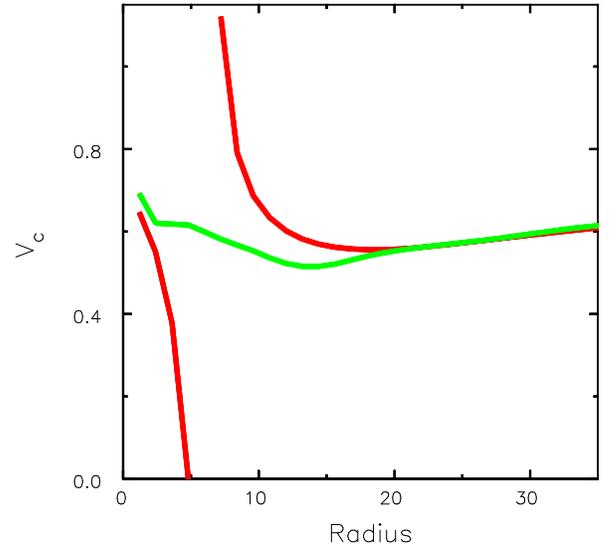}
\end{center}
\caption{The radial variation of circular speed at the final moment in
  models D (green) and E (red).  Despite having used the same initial
  disk and accretion rules, the resulting final rotation curves are
  remarkably different; being nearly flat in model D and strongly
  peaked, with a discontinuity in model E, where the attraction from
  the exterior massive annulus is outward.}
\label{fig:compDE}
\end{figure}

\begin{figure}
\begin{center}
\includegraphics[width=.9\hsize]{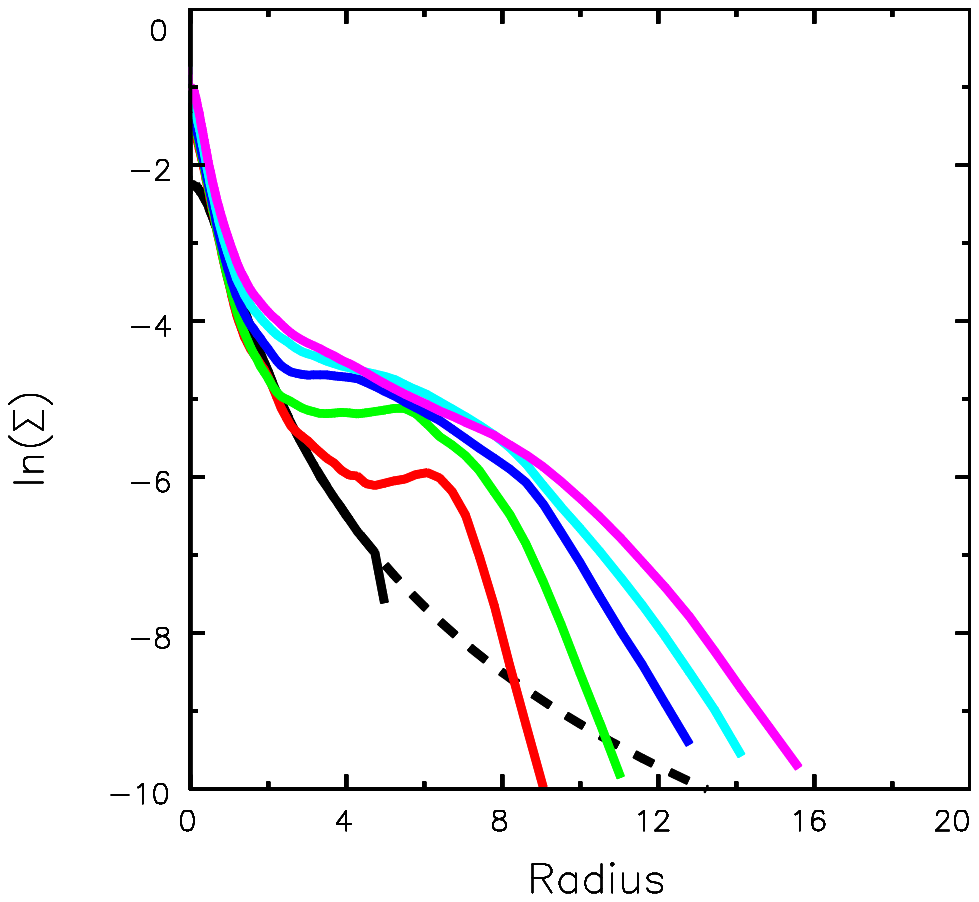}

\medskip
\includegraphics[width=.9\hsize]{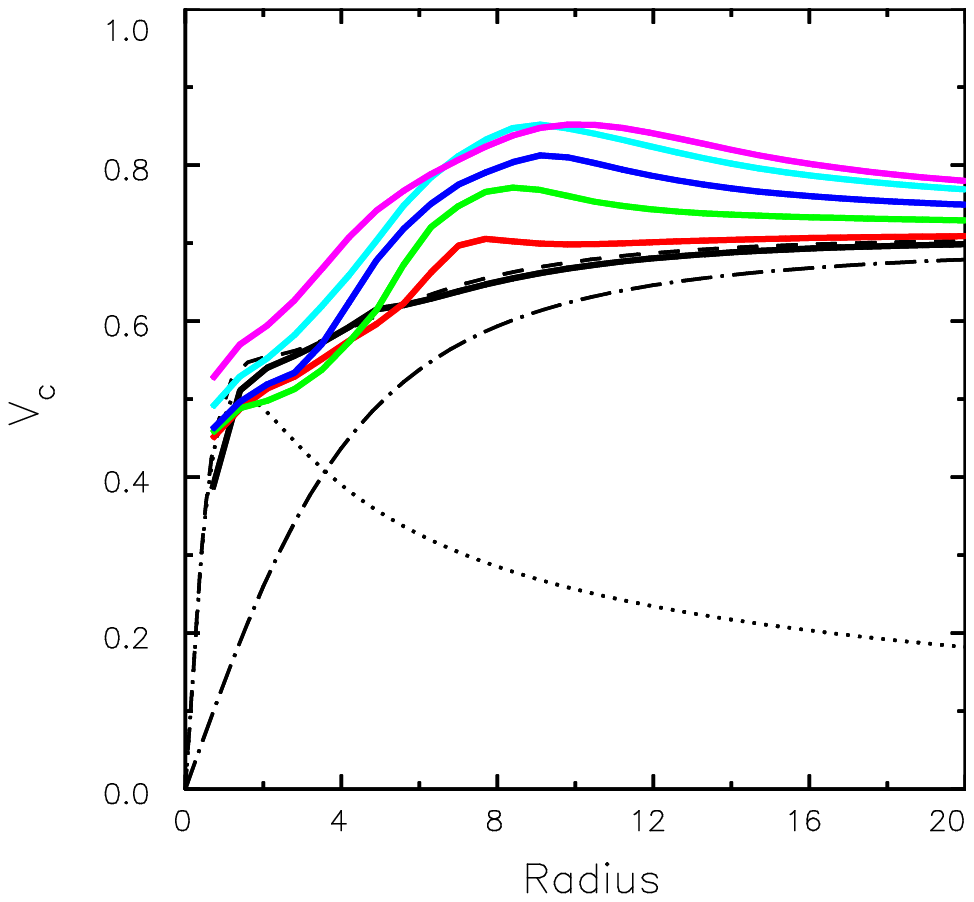}
\end{center}
\caption{The evolving surface density and circular velocity of a
  sample bulgeless disk. This is model F in Table~\ref{table:sims}.
  The initial conditions are the black line, while the red, green,
  blue, cyan, and magenta lines represent the state of the disk at
  $1000 \tau_0$ intervals. \emph{Upper:} The surface density of the
  disk from its initial state to $5000 \tau_0$. \emph{Lower:} The
  rotation curve produced by this system across the same time
  interval. As can be seen from both panels the activity in the galaxy
  is smoothing the features in the rotation curve and surface density.
}
\label{fig:318}
\end{figure}

\subsection{Models D and E}
The mean radius of added particles stayed constant in models A and B,
while we increased it in three steps in model C in a crude effort to 
mimic inside-out disk growth.  Here we experiment with outside-in
growth, by shifting the mean radius of the accretion annulus inwards
over time.  The details of the accretion rates and annuli used are
given in Table~\ref{table:sims} and mean radius of the annulus is
shifted inwards every $\sim 1560 \tau_0$.

Model D included, as in previous simulations, low-order
non-axisymmetric terms ($2 \leq m \leq 8$) in the forces acting on the
particles.  The evolution is illustrated in the usual manner in
Figure~\ref{fig:316}.  Here we again found that the spirals
redistributed matter, and the particles migrated radially, smoothing
out the initial rotation curve.  This model also developed a bar at
later times that was rather larger that in other cases, perhaps due to
the lower angular momentum material that was added later in the
simulation.  The final surface density profile, magenta line, has a
feature at $\sim 12a$, which is the approximate semi-major axis of the
bar.

The blue line shows that the density and rotation curve lacked this
feature at a time $2000\tau_0$ before the final moment.

The initial disk and accretion rules in model E were identical to
those in model D, except that the particles in model E experience
radial forces only; all sectoral harmonic terms with $m>0$ were
suppressed.  Since smoothing is caused by angular momentum changes driven
by collective non-axisymmetric disturbances, \ie\ spirals, the annulus
in model E cannot disperse because non-axisymmetric forces were
suppressed.  To emphasize the extent of mass rearrangement in model D,
we compare the circular speeds in models E (red) and D (green) at $t =
8000 \tau_0$ in Figure~\ref{fig:compDE}.  The circular speed in model
E actually becomes imaginary, \ie\ the radial attraction is
\emph{outwards} over a short radial range just interior to the massive
annulus, for reasons explained in \S1.2.  Because of the
unusual radial force law in model E, some particle orbits were
unstable and they acquired large radial velocities, whereas accreted
particles at other radii maintained near circular orbits.

\subsection{A bulgeless disk} \label{bulgless} 

\begin{figure}
\begin{center}
\includegraphics[width=.9\hsize]{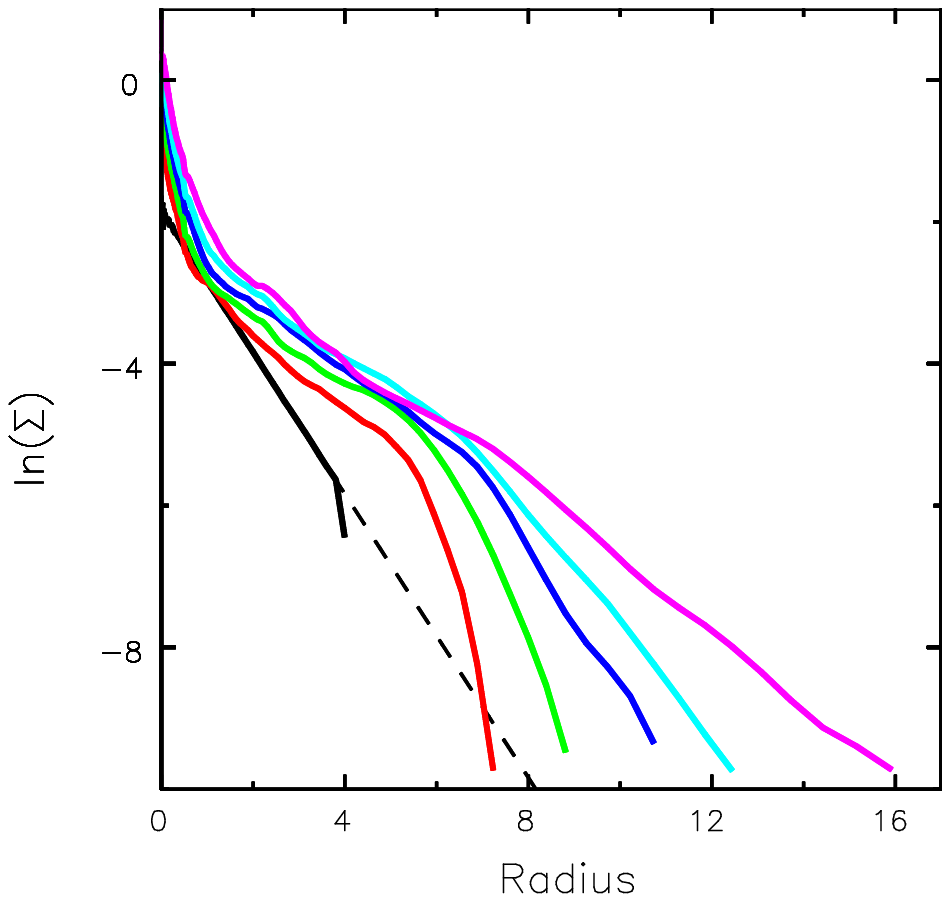}

\medskip
\includegraphics[width=.9\hsize]{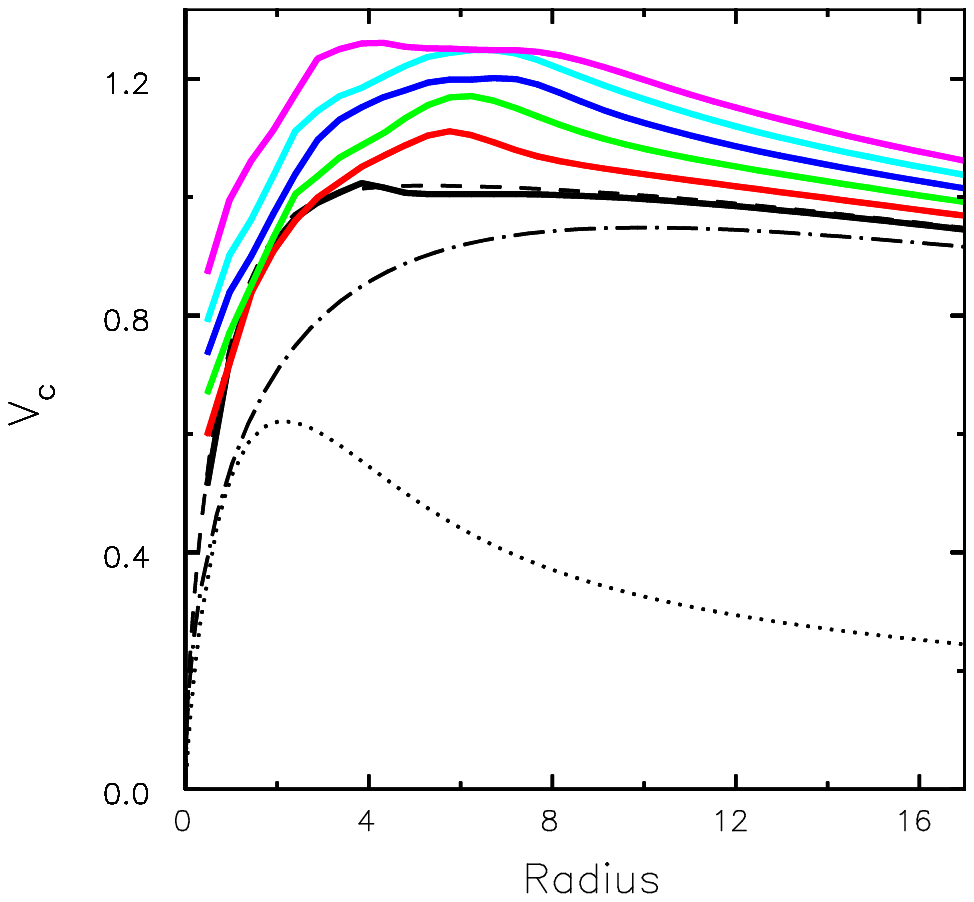}
\end{center}
\caption{The evolving surface density and circular velocity of a
  sample submaximal disk. This is model G in Table~\ref{table:sims}.
  The initial conditions are the black line, while the red, green,
  blue, cyan, and magenta lines represent the state of the disk at
  $1600 \tau_0$ intervals. \emph{Upper:} The surface density of the
  disk from its initial state to $8000 \tau_0$. \emph{Lower:} The
  rotation curve produced by this system across the same time
  interval. As can be seen from both panels the activity in the galaxy
  is smoothing the features in the rotation curve and surface
  density. The final rotation curve is flat over a range from $\sim
  3-8a$.  }
\label{fig:328}
\vspace{0.2cm}
\end{figure}

Thus far we have examined only cases with dense bulges that dominate
the central attraction near the center.  Model F has no bulge, but has
a halo core radius of $c = 5a$, smaller than in previous models.
Details of the mass components and accretion rule used in this case
are given in Table~\ref{table:sims}.

The results of adding material to the disk at large radius are
illustrated in Figure~\ref{fig:318}.  The feature in the rotation
curve caused by the accreted matter is far from erased in this case,
although matter has been redistributed somewhat.

Because of the dominant halo in the outer parts, the rotation curve
rises to $R =10a$ before beginning to decrease.  In many of our
simulations the dominant spiral modes are $m=2$ and $m=3$, but in this
case we see mostly $m=4$ in the outer parts where the halo dominates.
Low-mass disks support more multi-arm patterns (\S\ref{sec:maxdisk})
that are less effective at redistributing mass (see
\S\ref{sec:Discussion}).

\newpage
\subsection{A submaximal disk} \label{submax}

We turn now to examine a submaximal model, also with no bulge. Model G
uses an exponential disk with a massive rigid Hernquist halo. Details
of the mass components and accretion rule are provided in
Table~\ref{table:sims}.

The results of adding material to this disk may be seen in
Figure~\ref{fig:328}. The top panel of the figure illustrates the
surface density profile at $1600 \tau_0$ intervals. The initial
exponential disk grows from its initial length of $4a$ to $\sim 16a$
as the disk mass increases five fold.  The lower panel illustrates the
rotation curve of the disk at the same time intervals as the upper
panel.  Here it can be seen that the rotation curve has an initial
extra bump that is eventually flattened as the magnitude of the curve
rises.  The final (magenta) line rises in the innermost region, then
flattens over the range $3 \la R/a \la 8$, before decreasing.

\section{Smoothing mechanism} \label{sec:mestel}

While it is clear that spiral activity is the main agent responsible
for smoothing out features in the surface density profile and rotation
curve, we here try to achieve a deeper understanding of how this
happens.  To this end, we start with a smooth disk that is linearly
stable, and study the instabilities that are provoked by an accreted
annulus of additional matter.

\subsection{Mestel disk}
The Mestel disk is the only known linearly stable disk with
differential rotation.  This scale-free disk has the archetypal flat
rotation curve, with constant circular speed, $V_f$, at all radii and
the surface density
\begin{equation}
\Sigma(R) = \frac{q V_f^2}{2 \pi G R}
\label{eq.mestel}
\end{equation}
where $q$ is the active disk mass fraction.  \citet{Zang76} showed
that the full-mass disk, $q=1$, with an inner cut-out to break the
scale free character, and $Q \geq 1$, was stable to modes with $m=0$
and $m\ge2$, but was unstable to lop-sided modes with $m=1$.
\citet{Toom81} showed that the lop-sided mode could be suppressed by
reducing the active disk mass to one-half ($q=0.5$) while invoking a
rigid halo to maintain the same central attraction.  His linear
stability analysis predicted that such a disk with $Q=1.5$ would be
stable to \emph{all} small-amplitude non-axisymmetric modes.
Simulations by \citet{Sell12} seemed generally consistent with this
prediction, even though he found that non-linear changes, caused by
amplified shot noise from a finite number of particles, could lead to
the ultimate appearance of large amplitude non-axisymmetric
disturbances.  However, his simulations with moderately large $N$ did
not develop strong spiral features for a period long enough for the
disk to be considered stable our purposes.

\subsection{Simulations}
We therefore employ this model, with the inner cut-out and outer taper
described in \citet{Sell12}, but with $q = 1/3$ in order to
inhibit the late formation of bars through non-linear trapping by the
large-amplitude spirals we wish to provoke.  The central cut-out is a
moderately gentle taper in $L_z$ centered on a value $R_iV_f$, whose
(small) radius defines our length scale.  The outer taper is centered
on $R = 12R_i$, so that the disk has the surface density close to that
given by eq.~(\ref{eq.mestel}), with $q=1/3$, over the range $2 \la
R/R_i \la 9$.

In model M, we study the immediate response to the accretion of new
material and the effects of spiral activity on the matter distribution
over a short time scale.  The initial Mestel disk had $10^5$ particles
with $Q=1.5$ and we added particles at a rate of $10$ per time step to
the initial disk in an annulus of uniform density at $7R_i\pm2R_i$ for
just the first $2\,000$ time steps, thereby increasing the total
particle number by 20\% over a period of $\sim 1.14$ rotations of the
disk at the mean radius.

\begin{figure}
\begin{center}
\includegraphics[width=.9\hsize]{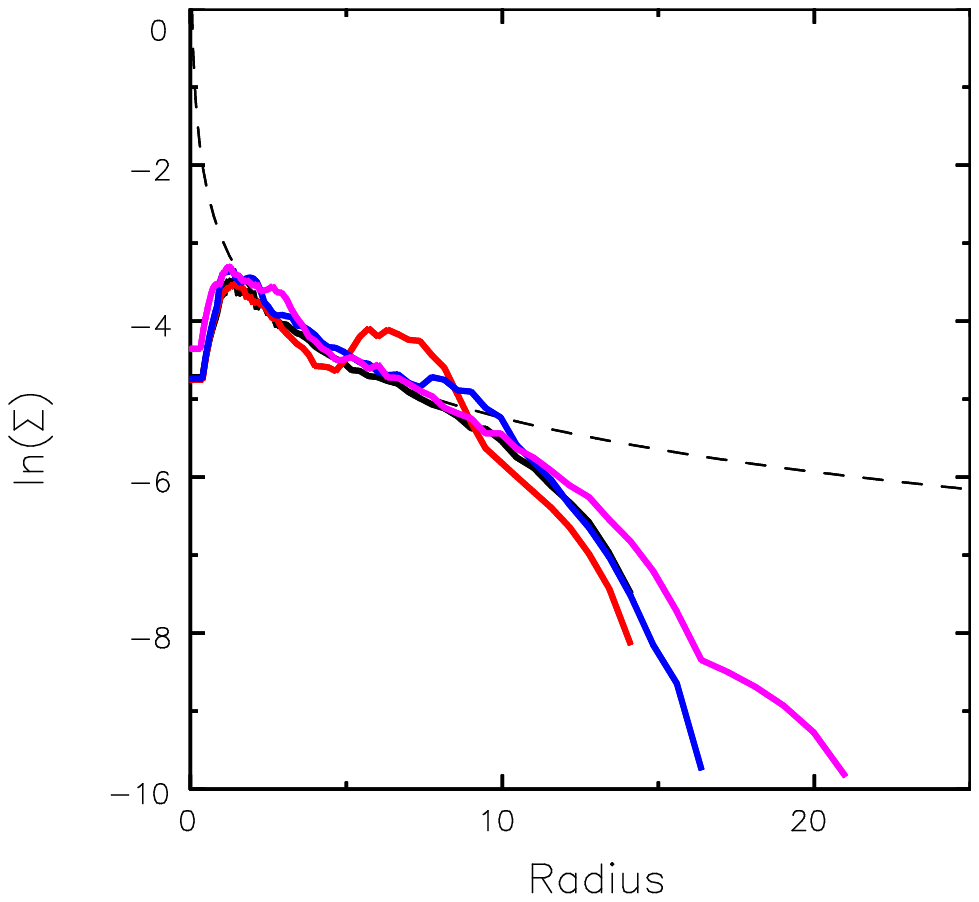}

\medskip
\includegraphics[width=.9\hsize]{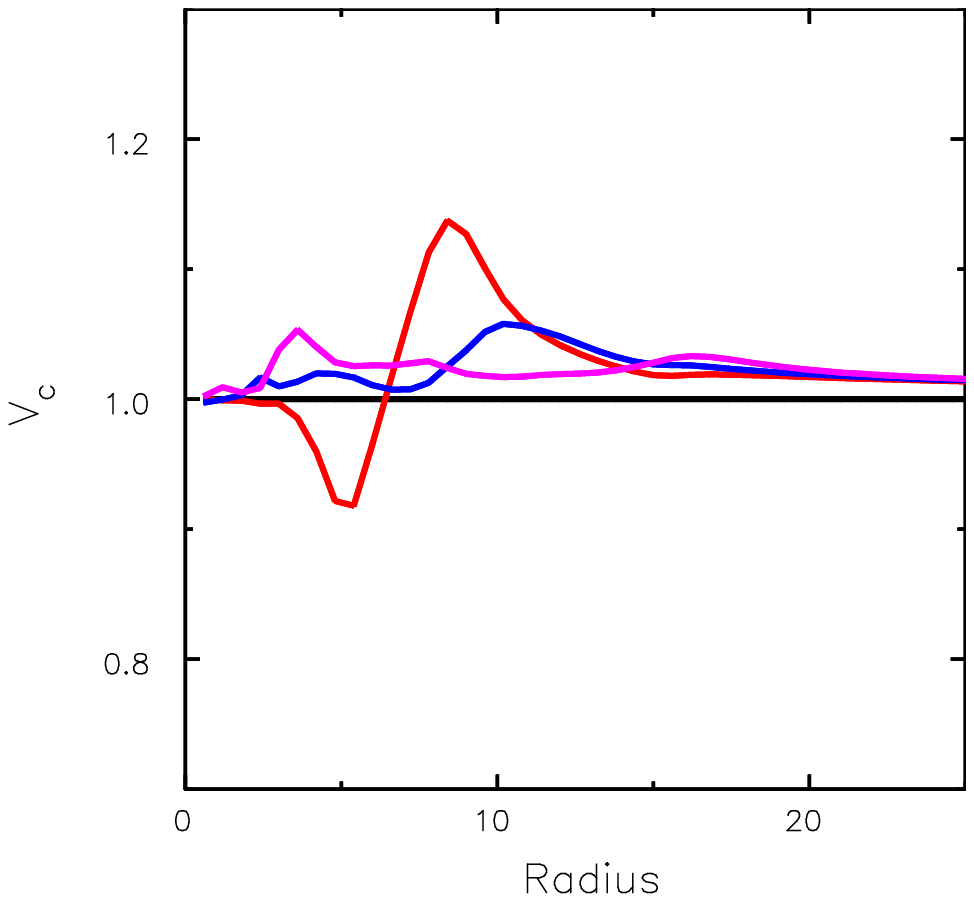}
\end{center}
\caption{The evolution of the surface density and the rotation curve
  for model M.  \emph {Upper:} The surface density of the disk
  initially (black), at $50$ $\tau_0$ (red), $100$ $\tau_0$ (blue),
  and at $250$ $\tau_0$ (magenta).  The ridge feature in the red curve
  is due to the particles added over the first $50$ $\tau_0$, and is
  smoothed out by spiral activity over the next five disk rotations.
  \emph{Lower:} The rotation curve of this model at the same interval
  as the upper panel.  The initial ridge feature leads to the
  pronounced feature in the rotation curve, which is also erased over
  the next five disk rotations as the matter is redistributed. }
\label{fig:248}
\end{figure}

\begin{figure*}
\begin{center}
\includegraphics[width=.23\hsize]{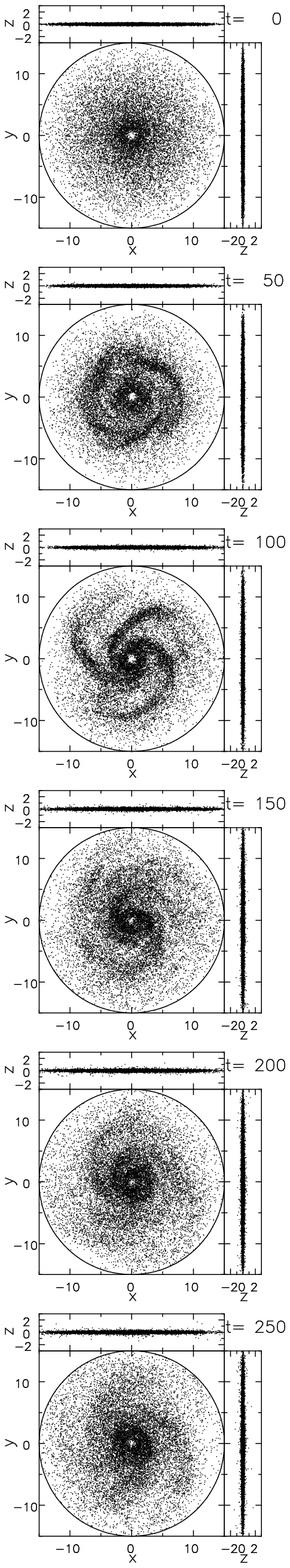}
\hfil
\includegraphics[width=.23\hsize]{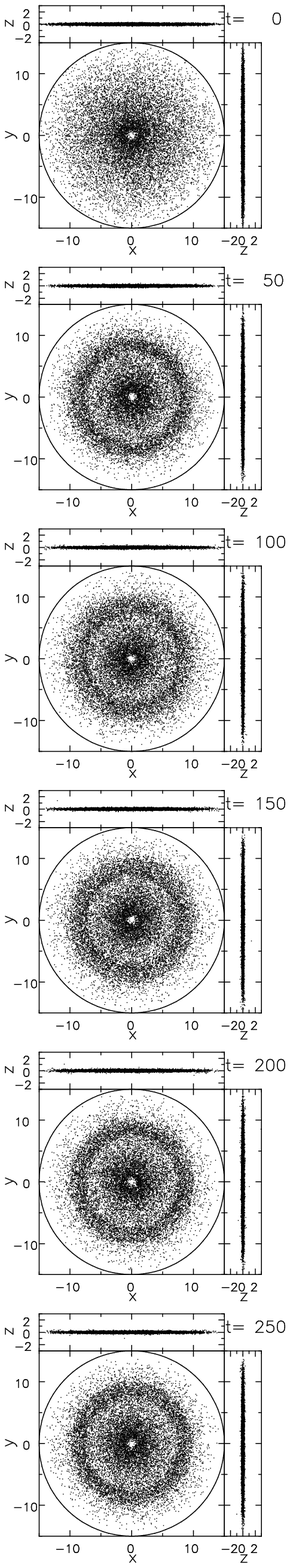}
\end{center}
\caption{The evolution of the Mestel disk simulations at the indicated
  times.  \emph{Left} Model M shows strong 3-arm spiral activity while
  the ring survives when spiral activity is suppressed in model N
  (\emph{right}). }
\label{fig:248-255}
\end{figure*}

The evolution of the surface density and circular speed in the
simulation are illustrated in Figure~\ref{fig:248}.  The dashed line
in the upper panel represents the initial profile of the untapered
disk, and the red curve shows the ridged density profile just after
the new particles have been added.  Subsequent evolution causes the
surface density profile to return to its original smoother state after
a just a few disk rotations at the mean radius of the added matter.

The radial variation of the circular speed is shown by the colored
lines in the lower panel at the same moments as shown in the upper
panel.  The addition of a substantial ring of extra mass centered on
$R=7$ gives rise to the inflexion, shown in red, in the originally
flat rotation curve (black).  The inflexion in the total circular
speed is caused by the central attraction of an annular disk of
matter, which is radially outward near the inner edge of the feature,
and strongly inward near the outer edge, as noted in \S1.2.

Although the disk without the annulus of accreted matter is stable,
the extra mass quickly provokes strong spiral patterns that cause mass
to be redistributed radially.  After just a few disk rotations, the
pronounced features in both the rotation curve and the surface density
profile are quickly smoothed, as shown by the blue and magenta lines.

To underscore that smoothing is caused by spiral activity, we have
rerun the simulation as model N with the same initial disk and
accretion rule, but with all non-axisymmetric sectoral harmonics
($m>0$) suppressed.  Figure~\ref{fig:248-255} compares the evolution
of model M, on the left, where strong 3-arm spirals appear, with model
N on the right, where spiral instabilities were artificially
suppressed and the added ring of particles was not dispersed.

\begin{figure}
\begin{center}
\includegraphics[width=.9\hsize]{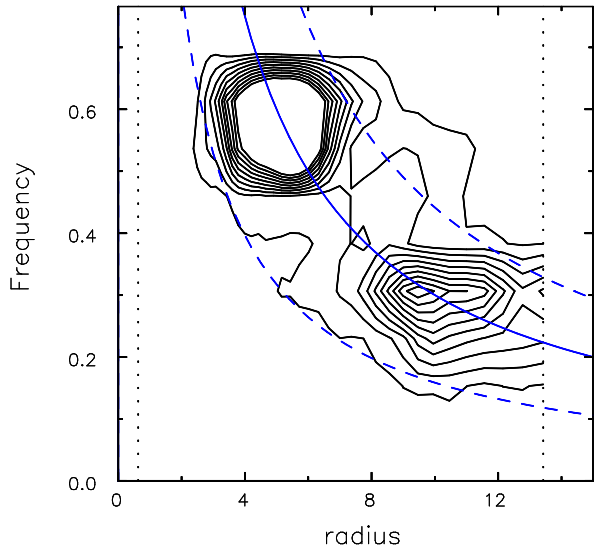}
\end{center}
\caption{Contours of power as functions of radius and frequency for
  the $m=3$ sectoral harmonics in our Mestel disk model.  Two modes
  develop after particles are added to the initial disk. }
\label{fig:spct248}
\end{figure}

\subsection{Spiral modes}
\citet{Sellwood&Kahn1991} studied instabilities provoked by both
grooves and ridges in the density distribution of similar models, and
also presented a local analysis of the normal modes of such features.
They found that an axisymmetric ridge possesses two linear modes for
each sectoral harmonic, but only those that could elicit a vigorous
supporting response from the surrounding disk would be unstable.  In
their local analysis, the two modes were expected to grow equally
rapidly, and to have corotation radii at some distance from the ridge
and symmetrically placed on either side of it.  Naturally, the
symmetry between the inner and outer modes, which is a consequence of
the neglect of curvature in their local analysis, is broken in a
global calculation.

\begin{figure}
\begin{center}
\includegraphics[width=.9\hsize]{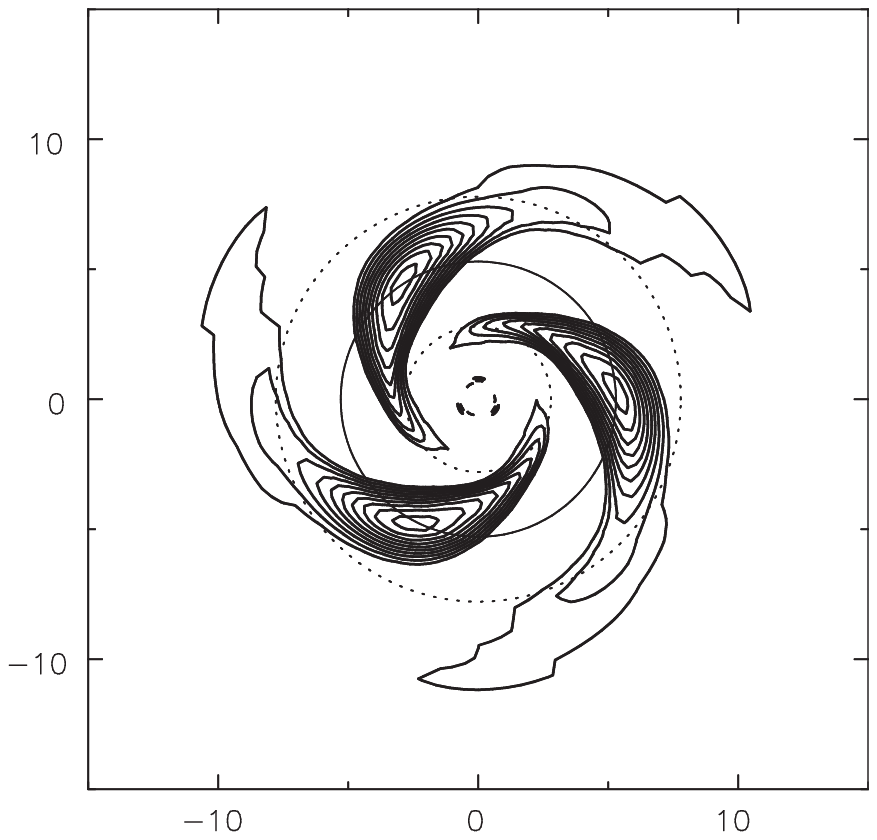}

\medskip
\includegraphics[width=.9\hsize]{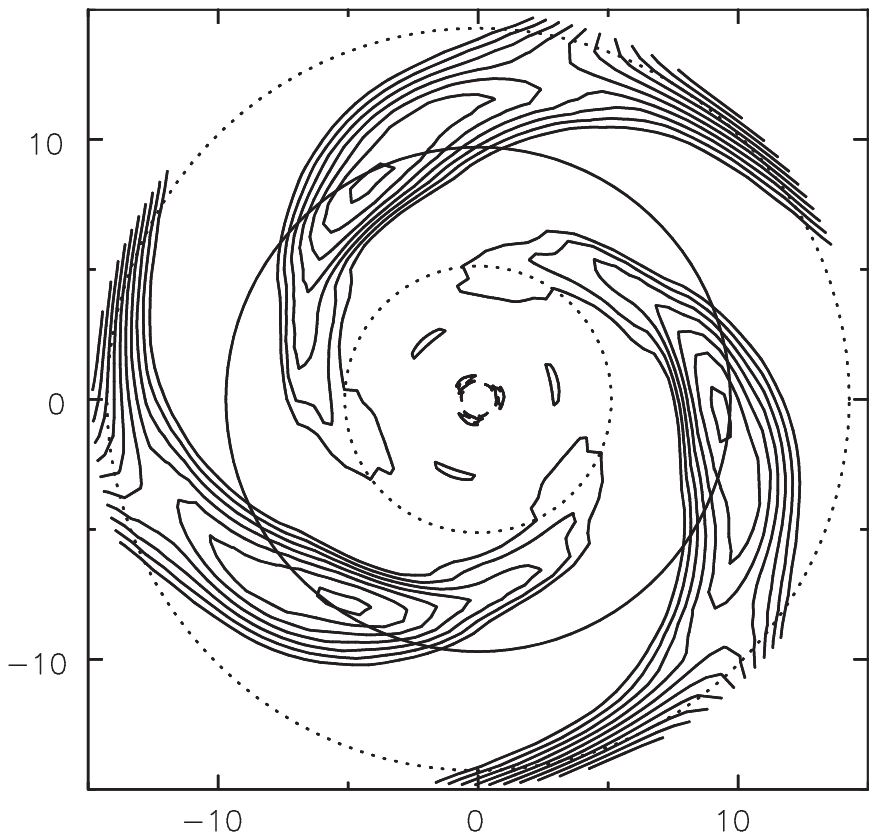}
\end{center}
\caption{Projected shapes of the two modes observed in the power
  spectrum of the Mestel disk. The contours illustrate the overdensity
  in the disk. The solid circles mark the corotation radius of the
  mode being illustrated. Similarly, the dotted circles mark the
  Lindblad resonances.  The corotation radii of the two modes sit on
  either side of the annulus of added particles. }
\label{fig:modes}
\end{figure}

The dominant unstable modes excited by the ridge should be those that
gain the strongest supporting response from the nearby disk.  The disk
supporting response is determined by swing-amplification, the theory
of which \citep{Toom81, BT08} predicts that the strongest response in
a Mestel disk will be for $m = q^{-1}$.  Thus we expect, and indeed
find, two rapidly growing modes at $m=3$.  Figure~\ref{fig:spct248}
illustrates the power spectrum of $m=3$ disturbances over the period
$30 \leq t/\tau_o \leq 110$, while the modes were mostly growing.  The
contours of power are broadened in frequency by the rapid growth of
the modes, but the Figure clearly shows two distinct disturbances with
frequencies $m\Omega_p \sim 0.3$ and $\sim 0.6$, and corotation radii
near $5R_i$ and $9R_i$, nicely straddling the mean radius of the added
matter.  The outer mode is clearly weaker than the inner, largely
because the dynamical time scale is almost twice as long at the
greater radius, and therefore the growth rate of the mode must be
lower also.  We have been able to fit the instabilities using the
apparatus described in \citet{SA86}, finding pattern speeds $m\Omega_p
\simeq 0.57$ and 0.32, although with highly uncertain growth rates.
The overdense parts of both modes are contoured in
Figure~\ref{fig:modes}.

As these two instabilities reach large amplitude, each opens up a
region around its own corotation radius where non-linear horseshoe
orbits appear \citep{SB02, BT08}.  This behavior arises because
particles orbiting at circular frequencies close to that of the
disturbance experience large changes in their angular momenta that
cause the home radii of their orbits to change with no increase in
random motion.  The sign of the change is to cause them to cross
corotation, \ie\ to reverse direction in the frame corotating with the
disturbance, hence the description ``horseshoe'' orbit.

Under normal circumstances, where the disk surface density declines
slowly outwards, the number of gainers of angular momentum is
approximately equal to the number of losers; particles simply exchange
places leaving the disk density profile pretty much as it was
\citep{SB02, SSS12}.  Here, however, corotation of each of the modes
is not far from the edge of the annulus, and the horseshoe region
extends into the higher surface density part of the disk, causing a
large excess of particles scattered out of that region over those that
are deflected into it -- in other words, the non-linear evolution of
both unstable modes rearranges matter in the sense of flattening out
the local density excess.  This is precisely the behavior we observe,
and is the fundamental reason that ridges and other features in the
surface density are smoothed by spiral activity.

\begin{figure}
\begin{center}
\includegraphics[width=.9\hsize]{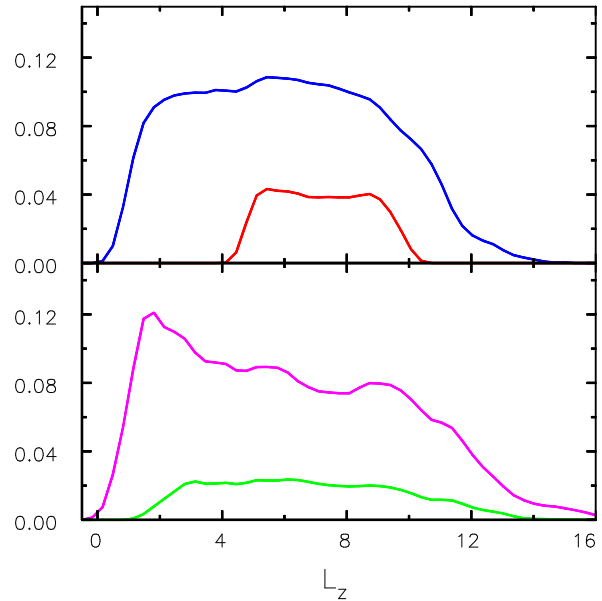}
\end{center}
\caption{The distribution of $L_z$ among the particles in model M, the
  linear scale on the $y$-axis is arbitrary.  \emph {Upper:} The blue
  curve is initial $L_z$ for the particles of the original disk and
  the red curve is the initial $L_z$ of the particles added over the
  run.  \emph{Lower:} The final angular momenta (at $t=250\tau_0$) of
  the original disk particles (magenta) and of the added particles
  (green).  Note the dramatic change to the distribution of the added
  particles by the end.}
\label{fig:lzpcs}
\end{figure}

\begin{figure}
\begin{center}
\includegraphics[width=.9\hsize]{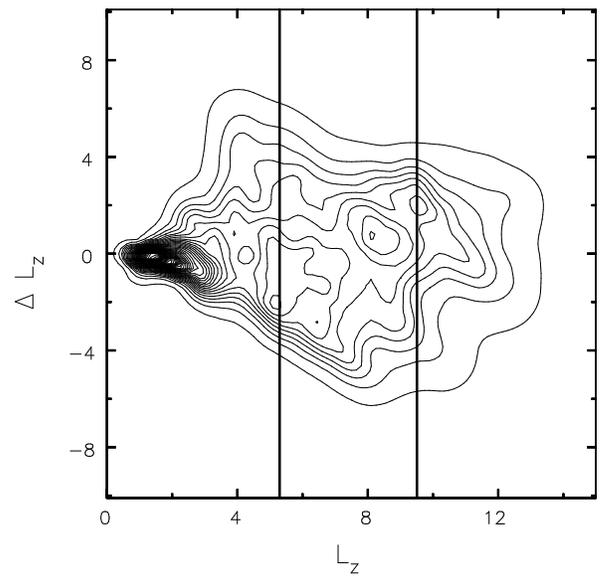}
\end{center}
\caption{Contours show the density of all the particles in the space
  of their initial $L_z$ and $\Delta L_z$ by time $250\tau_0$ in model
  M.  These changes are dominated by the two transient spiral modes
  acting on the particles, with corotation at the marked values of
  $L_z$.  The vertical lines mark the $L_z$ values of corotation for
  the two modes.}
\label{fig:lzdiff}
\end{figure}

\subsection{Angular momentum changes}
The distributions of angular momenta of the original disk particles
and of the added particles are shown separately in
Figure~\ref{fig:lzpcs}; the initial values in the upper panel and
those at the end of the simulation ($t= 250\tau_0$) in the lower.  The
initial distribution of the added particles, which were placed on
circular orbits, reflects the radial extent of the annulus where they
were placed.  The distribution of $L_z$ values of the added added
particles at the later time (lower panel) has spread quite remarkably,
while that of the initial disk particles was changed to a lesser
extent.

Figure~\ref{fig:lzdiff} shows density contours in the $L_z-\Delta L_z$
plane.  The ridges with negative slope reflect the large changes
brought about by horseshoe orbits at corotation, where the gainers
were inside corotation and the losers outside.  The two vertical lines
mark the angular momenta at corotation for the modes we identified;
most of the changes can be attributed to these two modes, but some
further minor changes occurred at smaller $L_z$ that were probably
caused by secondary instabilities.

\subsection{Potential vorticity}
Although their paper has received little attention in the realm of
disk galaxy dynamics, \citet[][hereafter LH78]{Lovelace&Hohlfeld1978}
predicted instabilities associated with radially localized density
enhancements.  Their analysis focused on the radial gradient of the
quantity known as the potential vorticity, or sometimes
``vortensity,''
\begin{equation}
f(R) = {\Sigma(R) \over \omega_z},
\end{equation}
where $\omega_z$ is the $z$-component of vorticity.  (For a laminar,
axisymmetric flow on circular orbits, $\omega_z = (1/R)dL_z/dR =
\kappa^2 / 2 \Omega$.)  Their local fluid analysis revealed that
instabilities should arise wherever there were steep radial variations
in $f$, and they focused on models without a large-scale gradient but
having both ridges and grooves in this quantity \citep[as did][for
  stellar dynamical models]{Sellwood&Kahn1991}.  LH78 noted that in a
massive disk having a flat rotation curve, \ie\ the Mestel disk, all
three quantities $\Sigma$, $\Omega$, and $\kappa$ vary inversely with
$R$, and therefore $f$ has no radial gradient and the disk should be
locally stable by their criterion.

\begin{figure}
\begin{center}
\includegraphics[width=.9\hsize]{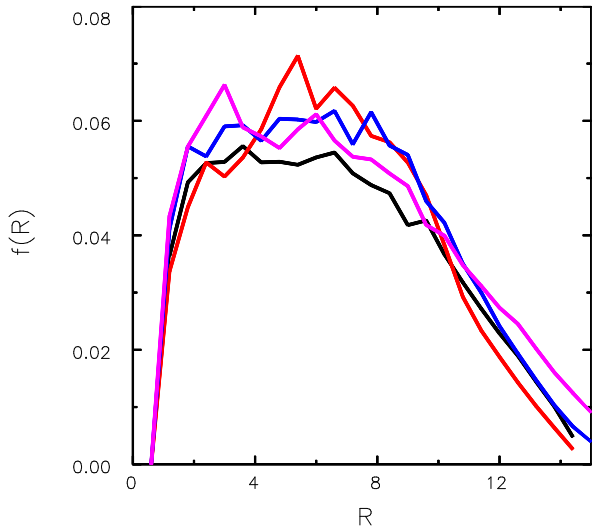}
\end{center}
\caption{The evolution of the potential vorticity, $f =
  2\Sigma\Omega/\kappa^2$, in model G.  The line colors have the same
  meaning as in Fig.~\ref{fig:248}.  See the text for a
  discussion.}
\label{fig:248v}
\end{figure}

A ridge, or a groove, should be destabilizing, and LH78 argued `an
important nonlinear effect of the fluctuations (from the resulting
instabilities) may be the radial transport of mass, angular momentum,
and energy. ... A possible mechanism of saturation of the linear  
instability is that the ``induced'' radial transport acts to smooth
out the maxima and minima in $f(R)$ with the result that $f(R)
=\;$constant for at least a range of $R$.'  The behavior in our Mestel
disk simulation conforms closely to that anticipated in these
far-sighted remarks.

Figure~\ref{fig:248v} shows the radial variation of $f$ in model M at
the same times as the other quantities in Figure~\ref{fig:248}.  The
initial value (black) is flat over $2 \la R/R_i \la 9$, where the
surface density of our modified Mestel disk is little affected by the
inner and outer tapers.  The accreted annulus introduced a pronounced
spike at first (red), which was later smoothed (blue).  A new spike
appeared at smaller radius later (magenta) perhaps because the angular
momentum changes associated with flattening the annulus had to be
deposited somewhere.

We have examined the radial variation of $f$ in our other models,
which always manifested declining profiles.  Bumps in the broadly
declining profiles, created by our accretion rules, were generally
removed, however.  These models differed from the Mestel disk in many
ways, especially because the central attractions from neither the
inner bulge nor the outer halo were self-similar, whereas that was
true for the Mestel disk.  A tendency to flatten was marginally more
convincing in Run F, which lacked a central bulge, than in runs A
through D.

\begin{figure}
\begin{center}
\includegraphics[width=\hsize]{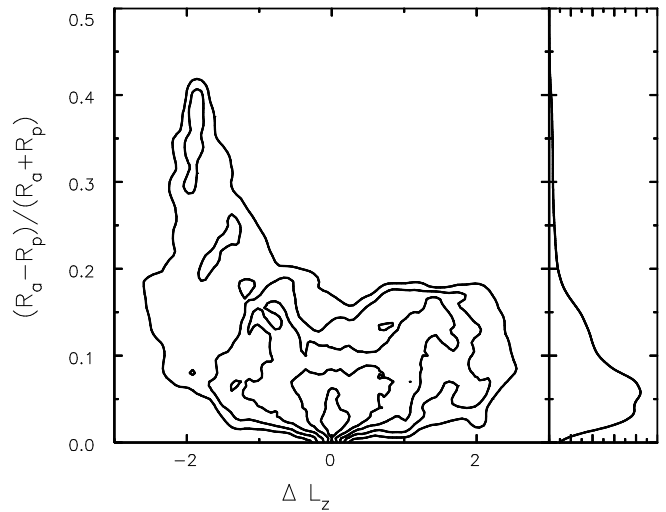}
\end{center}
\caption{The density of recently added particles as functions of their
  change in angular momentum, $\Delta L_z$, and eccentricity,
  $\epsilon = (R_a-R_p)/(R_a+R_p)$, at $t=2000\tau_0$ in model D.  The
  contours are logarithmically spaced in order to reveal the low
  density regions.  The RH panel, which shows the density distribution
  on a linear scale in arbitrary units projected onto the y-axis,
  reveals that the orbits of most particles have remained close to
  circular, with $\epsilon<0.1$, while they experienced up to 50\%
  changes in $L_Z$.  The very few particles that lost $L_z$ were
  scattered to more eccentric orbits.}
\label{fig:Lzecc}
\end{figure}

\section{Radial migration}
We have just shown that changes to the home radii of particles,
\ie\ radial migration, smoothed the rotation curve and mass profile of
the Mestel disk.  We here show that this behavior also played the
major role in smoothing the mass distributions in our other models.  A
painstaking demonstration of how spiral disturbances drove the orbital
evolution of even a fraction of the particles in our simulations
\citep[\eg][]{Roskar12} would be best presented in a separate paper.
So here we simply demonstrate that particles do not gain much random
energy as their angular momenta are changed, which is a characteristic
feature of radial migration driven by scattering at corotation.

We therefore examined the orbital properties of the $64\,000$
particles that were added in a narrow annulus to model D over the time
interval $1900 \leq t/\tau_0 \leq 2000$.  As a measure of random
motion, we computed the eccentricity $\epsilon = (R_a-R_p)/(R_a+R_p)$
of the particle orbits in the frozen, axisymmetric potential of the
model at $t = 2000$.  Here $R_p$ and $R_a$ are the peri- and
apo-centric radii of each particle in the disk mid-plane.  In fact,
the simulation supported strong spiral patterns at all times,
including this moment, so our procedure of computing the eccentricity
in an axially symmetrized potential causes $\epsilon$ to be
overestimated, because it treats the non-axisymmetric spiral streaming
motions as random motion.

Figure~\ref{fig:Lzecc} shows the density of all the recently added
particles as functions of their angular momentum change since they
were added to the simulation, $\Delta L_z$, and eccentricity
$\epsilon$ at $t=2000$.  The added particles had initial $L_{z,i}
\simeq 3.6$, while $|\Delta L_z|$ in Figure~\ref{fig:Lzecc} ranges up
to 2, in our units.  A small fraction of angular momentum losers lie
in a tail reaching to $\epsilon \simeq 0.4$, which are likely to have
been scattered at an ILR, as observed by
\citet{Sellwood&Carlberg2014}.  Most of the particles for which
$|\Delta L_z| \la 1$, or a 25\% change, have remained on nearly
circular orbits, with $\epsilon < 0.1$.  This is strong evidence for
radial migration driven by surfing near the corotation resonances of
spiral waves.

The total angular momentum changes over a longer period in a different
run, presented in Fig.~\ref{fig:307dLz}, imply that some particles
have lost almost all their angular momenta, while others have gained
more than twice their original.  Since a strong spiral can cause
$\Delta L_z \la 20\%$, these much greater changes must have resulted
from multiple spiral events.

Angular momentum changes are caused by the gravitational torque, which
varies directly with the spiral potential amplitude \citep[][Appendix
  J]{BT08}.  The torque is greatest for strong, open patterns with 2-
and 3-fold rotational symmetry.  It is weaker for multi-arm patterns
for two reasons.  (1) For a given density amplitude, the potential
variations of waves that vary on smaller spatial scales, i.e.\ as $m$
rises, are weaker, and (2) the amplitudes of the spirals that develop
in our models F and G were also lower.  To quantify the second point,
we define $\delta_m = \Sigma_m/\Sigma_0$, from a Fourier decomposition
of the surface density at any radius, and find $\langle \delta_2
\rangle$ in run A is about seven times greater than $\langle \delta_4
\rangle$ in run G, where the angle brackets denote averages over a
broad swath of radii and time.  Thus spiral torques are substantially
weaker for our multi-arm patterns.  Spirals are also weaker in disks
having a higher $Q$, which is a direct measure of the competition
between the growth of features due to gravitationally-driven
collective effects and their dispersal due to random motion.

Thus the spirals that smooth the density profile most effectively have
2- or 3-arms.  Such patterns are preferred in cool, responsive disks
with high surface density for reasons given in \S\ref{sec:maxdisk}.
Clearly, therefore, features in the disk density and rotation curve
are smoothed out more readily when the disk is heavy, and spirals have
much less of a smoothing effect in submaximal disks, as we have found.

\section{Discussion} \label{sec:Discussion}

Our experiments have shown that angular momentum changes caused by
spiral instabilities in near-maximal disks smooth out both
irregularities in the mass distribution within the disk and the
rotation curve.  Our experiments were artificial in many respects, but
we believe that the mechanism they have revealed is robust.

As we described in \S\ref{sec:galform}, the disks of real galaxies are
thought to have grown through the addition of gas that settles onto
circular orbits in the disk mid-plane as it cools and forms stars.
The radius at which gas settles is determined by both its angular
momentum and the central attraction of the dominant mass components.
The purpose of our study was to determine how sensitive is the final
disk mass profile and rotation curve to the distribution of angular
momentum in the accreted matter.  We therefore needed to be able to
vary this at will.  Were we to have included gas in our
models, we would have had to vary its angular momentum distribution
and cooling rate in a similarly unrealistic manner to have achieved
our science goal.  We therefore opted for the simplest imaginable
accretion formula to enable us to explore the widest possible range of
rules in the most efficient and direct manner.

We employed rigid halos and bulges for similar reasons.  In reality,
responsive spheroidal components must compress as the disk grows and
the central attraction strengthens.  Also, some angular momentum
exchange between these components and the disk may well occur.  By
keeping them rigid and unresponsive, we have been able to isolate the
smoothing effect of the spiral activity in the disk, which would have
been less clear if bulge and halo interactions were also taking place.

Halo compression has been invoked \citep{Burstein&Rubin1985} as a
possible reason for the absence of a feature in the rotation curve at
the radius where the central attraction changes from disk domination
to halo domination, which is the observed disk halo conspiracy
\citep{Bahcall&Casertano1985}.  Here we have shown that spiral
activity rearranges the disk matter to erase any feature, without help
from halo compression.

The accreted matter in all our simulations was placed in axisymmetric
rings, as opposed to non-axisymmetric distributions that may be
expected in a hierarchical cosmological context.  However, accretion
of a blob, or of a stream of cold gas over a narrow range in azimuth,
will only excite spiral responses more easily, again leading to rapid
spreading, as we confirmed in additional simulations not presented
here.

We added mass at a steady rate in many, but not all, of our
simulations, whereas hierarchical galaxy formation would suggest that
a quiescent accretion history may be unusual.  However, the behavior
in the Mestel disk case, \S\ref{sec:mestel}, showed that the smoothing
mechanism persists when matter is added for a short period only.

Since bar formation has long been known to cause a substantial
redistribution of angular momentum within a disk \citep{Hohl1971,
  Sell81, Deba04, Bere06}, we have tried to prevent it from happening
in our experiments, in order to isolate the effects of spirals.
However, bars have formed at a late stage in a number of our
simulations, but usually late enough that spiral activity dominated
most of the evolution.

The majority of our models are of heavy disks whose rotation curves
become largely flat or even declining (\eg\ models A through D).  The
final rotation curves in these cases are generally flat and about as
smooth as those of many galaxies.  Good examples for comparison with
these models are M31 \citep{Corbelli2010}, which has a broad hump over
the range $5 \la R \la 15\;$kpc where the disk makes its peak
contribution to rotation curve, and NGC~5055 \citep{deBlok08} which
has the feature noted by \citet{Sancisi2004} associated with the bulge,
and is not perfectly flat at larger radii.

The more dominant halo models F and G still manifested some spreading
of the accreted matter (see Figs.~\ref{fig:318} \& \ref{fig:328}), but
it was insufficient to flatten the rotation curves, which rose in the
inner parts right through to the end of our simulations.  Because the
disk was less dominant in both these cases over the region where mass
was added, the spiral patterns that developed in these two models had
higher multiplicities ($m \ga 4$) than in models A-D, where two-armed
patterns were the strongest.

While none of our models has resulted in a perfect exponential disk,
the surface density profiles at the final times, shown by the magenta
curves in Figures~\ref{fig:282}, \ref{fig:307}, and \ref{fig:305} are
much more nearly exponential than that of the original disk plus
accreted matter.  The smoothing mechanism we identify here is clearly
of importance to the origin of the classical exponential disk profile
\citep{Freeman1970}.  \Ignore{Indeed, deep photometry
  \citep{vanderKruit79, vanderKruit07, EPB08, Martin-Navarro14} has
  revealed departures from a perfect exponential in many galaxies,
  suggesting that an exponential profile over many scale lengths is
  quite rare.}

\section{Conclusions} 
\label{sec:Conclusions}

All our experiments with massive disks have shown that mass added to a
disk, or even exterior to it, having a narrow range of angular
momentum is quickly spread in radius by spiral patterns.  While the
overall size the disk must be determined by its total angular momentum
content, our results argue that the smoothness of the density profile
is remarkably insensitive to the distribution of angular momentum
among the accreted baryons.  Our accretion rules were deliberately
chosen to demonstrate the ability of spiral activity to lead to
physically reasonable disks from even quite unrealistic distributions
of accreted matter.  Disks are believed to grow in an inside-out
fashion, but we have shown that even outside-in growth leads to the
same behavior, and a similar outcome.

The accreted material at first settles to the radius at which its
angular momentum is in centrifugal balance.  We find that the observed
generally smooth rotation curves and decreasing surface density
profiles of disk galaxies are created by internally driven secular
processes, especially in disks that support strong spiral patterns
with low-order rotational symmetry.  Thus the final density profile is
insensitive to the detailed angular momentum distribution of the
infalling baryons, or to their distribution of arrival times, as they
settle into the halo potential well.

Angular momentum changes driven by spiral patterns are responsible for
smoothing out irregularities in the radial mass profile.  Our detailed
study of the instabilities of a single annulus added to an otherwise
stable disk (\S\ref{sec:mestel}), showed that the non-linear evolution
of the unstable modes excited by the density ridge disperses the
density excess through horseshoe orbits.  The excess of those
scattered out of the density ridge over those scattered into it by
these orbits spread out the ridge in less than $200\tau_0$, which is
$\sim 4$ disk rotation periods at the mean radius of the annulus.
Since horseshoe orbits cross corotation for the spiral mode, the stars
do not gain significant random motion \citep[none to first
  order,][]{SB02}, and therefore the only heating associated with the
smoothing process is caused by the much more minor angular momentum
changes at the Lindblad resonances \citep{Sellwood2014}.

We believe that the mechanism of mode excitation that we identified in
the Mestel disk holds more generally in our other experiments, as was
predicted by the local analysis by \citet{Lovelace&Hohlfeld1978} and
by \citet{Sellwood&Kahn1991}.  These modes erase the features that
excited them through angular momentum changes around their corotation
radii, which change the home radii of particles without increasing
random motion (see Fig.~\ref{fig:Lzecc}).  The radial extent of the
horseshoe region is limited to $\sim \pm 20$\% of the radius of
corotation for strong spiral patterns with few arms.  Thus multiple
spiral episodes with a variety of corotation radii are required to
spread the accreted material over the full radial extent of the disk.
The smoothing effect is weaker in disks that support weaker spirals
and/or patterns of higher higher multiplicity.  The dominant patterns
in large spiral galaxies do have 2- or 3-fold rotation symmetry
\citep[\eg][]{Davis2014}, but more multi-arm patterns are seen is some
dwarf galaxies, \eg\ NGC 3928 \citep{vdB80, Carollo97}.

Continuous accretion, which we applied in the majority of cases,
causes a succession of spiral instabilities associated with the
density excess, and probably also secondary instabilities caused by
other features in the density profile where particles have accumulated
after having been moved from the original ridge.  However, none of our
models has a perfectly smooth mass profile or rotation curve, and the
mild features in the one are reflected in similar mild features in the
other, just as for ``Renzo's rule'' \citep{Sancisi2004}.  This
mechanism also provides at least a partial explanation of the long
standing disk-halo ``conspiracy.''

\acknowledgments 

This work was supported by NSF grant AST/1108977.  We thank the
anonymous referee for a helpful report.

\bibliography{smooth}

\end{document}